\documentclass[prl,twocolumn]{revtex4-1}
\usepackage{amsmath}
\usepackage{amssymb}
\usepackage{graphicx}
\usepackage{esint}
\usepackage{bm}
\usepackage{braket}
\usepackage{hyperref}
\usepackage{color}

\definecolor{myblue}{RGB}{0,0,255}
\hypersetup{
    colorlinks,
    citecolor=myblue,
    linkcolor=myblue,
    urlcolor=myblue
}

\begin{document}
\title{Quantum Thermodynamic Uncertainty Relation for Continuous Measurement}
\author{Yoshihiko Hasegawa}
\email{hasegawa@biom.t.u-tokyo.ac.jp}
\affiliation{Department of Information and Communication Engineering, Graduate
School of Information Science and Technology, The University of Tokyo,
Tokyo 113-8656, Japan}

\date{\today}
\begin{abstract}
We use quantum estimation theory to derive a thermodynamic uncertainty relation in Markovian open quantum systems,
which bounds the fluctuation of continuous measurements. The derived quantum thermodynamic uncertainty relation holds for arbitrary continuous measurements satisfying a scaling condition.
We derive two relations; the first relation bounds the fluctuation
by the dynamical activity and the second one does so by the entropy production. 
We apply our bounds to a two-level atom driven by a laser field and a three-level quantum thermal machine with jump and diffusion measurements. 
Our result shows that there exists a universal bound upon the fluctuations, regardless of continuous measurements. 
\end{abstract}
\maketitle
\emph{Introduction.}---Uncertainty relations distinguish the possible from the impossible and have played fundamental roles in physics. 
Recently, thermodynamic uncertainty relations (TURs) have been found in stochastic thermodynamics, showing that the fluctuation of time-integrated observables is lower-bounded by thermodynamic costs, such as entropy production and dynamical activity \cite{Barato:2015:UncRel,Barato:2015:FanoBound,Gingrich:2016:TUP,Polettini:2016:TUP,Pietzonka:2016:Bound,Horowitz:2017:TUR,Proesmans:2017:TUR,Pietzonka:2017:FiniteTUR,Pigolotti:2017:EP,Garrahan:2017:TUR,Macieszczak:2018:TURLR,Dechant:2018:TUR,Barato:2018:PeriodicTUR,Koyuk:2018:PeriodicTUR,Vroylandt:2019:PowerTradeOff,Dechant:2018:FRI,Terlizzi:2019:KUR,Hasegawa:2019:CRI,Dechant:2019:MTUR,Ito:2018:TimeTUR,Hasegawa:2019:FTUR,Vu:2019:UTURPRE,Vu:2019:TURProtocol} (see \cite{Horowitz:2019:TURReview} for review). TURs predict the fundamental limit of biomolecular processes and thermal machines, and they have been applied to infer the entropy production \cite{Seifert:2019:InferenceReview,Li:2019:EPInference,Manikandan:2019:InferEP}. 

In contrast to classical systems, studies of TURs in the quantum regime are in very early stages. One of the distinguishing properties of quantum systems is how they behave under measurement. In stochastic thermodynamics, it is naturally assumed that we can measure the stochastic trajectories of the system. 
In quantum systems, output is obtained through measurements, but the measurements themselves alter the system state. Moreover, in addition to the freedom of how we compute the current in stochastic thermodynamics, we have an extra degree of freedom based on how we measure the quantum systems. Although TURs have been recently studied in quantum systems
\cite{Erker:2017:QClockTUR,Brandner:2018:Transport,Carollo:2019:QuantumLDP,Liu:2019:QTUR,Guarnieri:2019:QTURPRR},
these works have not considered the measurement effects explicitly, or specified a type of measurement in advance. 

In this Letter, we derive a quantum thermodynamic uncertainty relation (QTUR) for Markovian open quantum dynamics using quantum estimation theory \cite{Paris:2009:QFI,Liu:2019:QFisherReview,Gammelmark:2014:QCRB}. In Ref.~\cite{Hasegawa:2019:CRI}, we have derived a TUR for Langevin dynamics via the Cram\'er--Rao inequality. Extending this line of reasoning to quantum dynamics, we derive a QTUR for continuous measurements with the quantum Cram\'er--Rao inequality. The quantum Cram\'er--Rao inequality holds for arbitrary measurements, while the classical one is satisfied for specific measurements, indicating that the quantum version is more general. By virtue of this generality, the obtained QTUR holds for arbitrary continuous measurements satisfying a scaling condition (cf. Eq.~\eqref{eq:obs_scaling}).
Our QTUR has two variants; the first relation is bounded by the dynamical activity, and the second by the entropy production. We demonstrate the QTUR with a two-level atom and a quantum thermal machine under jump and diffusion measurements.

\begin{figure}
\includegraphics[width=8cm]{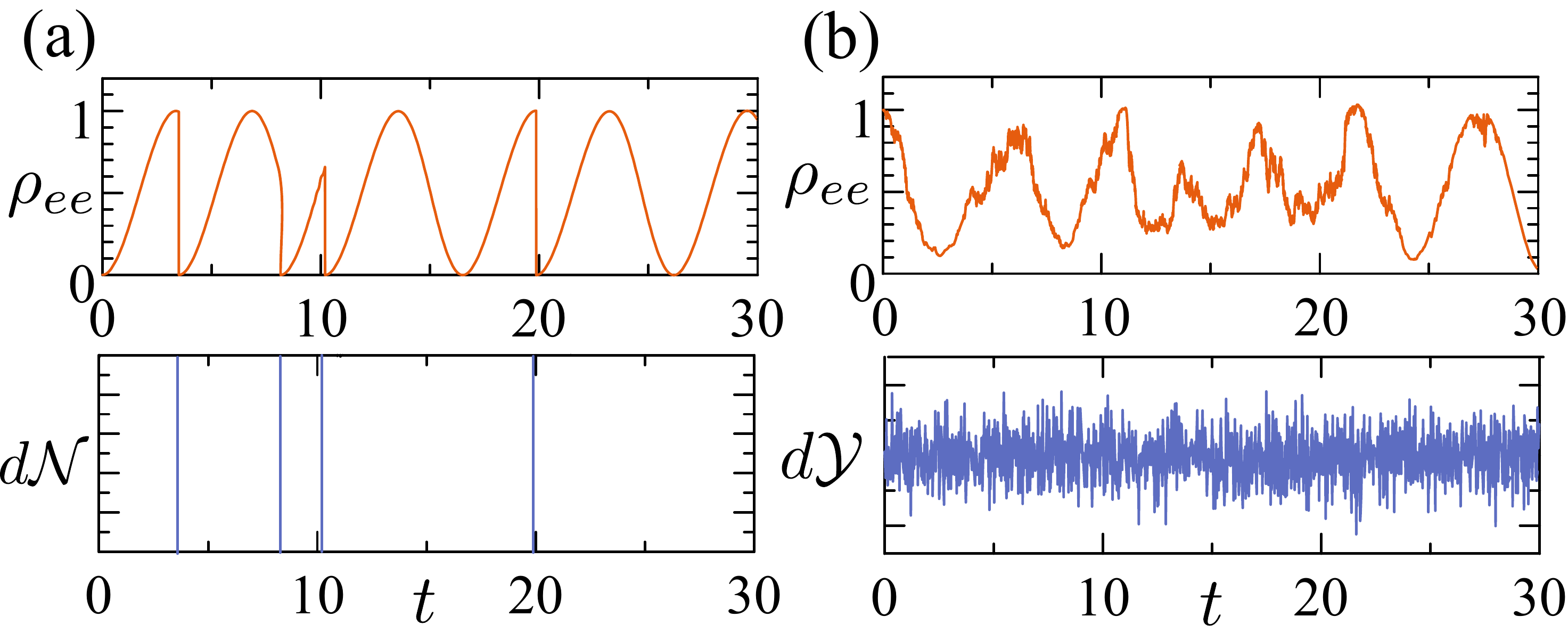}

\caption{Quantum trajectories and measurements of (a) jump measurement (photon counting) and (b)
diffusion measurement (homodyne detection) in a two-level atom. Upper panels are quantum
trajectories of $\rho_{ee}\equiv\braket{\epsilon_{e}|\rho|\epsilon_{e}}$
and lower panels are measurement outputs. \label{fig:method_ponch}}
\end{figure}

\emph{Methods.}---The TURs in classical stochastic thermodynamics consider the fluctuation of currents, which are time integrals of the stochastic trajectories. Analogously, we wish to bound the fluctuation of the time-integrals of continuous measurements in quantum dynamics. 

In continuous measurements, we consider a principal system $S$ and an environment $E$. Consider a Kraus operator $\mathcal{V}_m$ acting on the principal system, which satisfies $\sum_m \mathcal{V}_m^\dagger \mathcal{V}_m = \mathbb{I}$ ($\mathbb{I}$ denotes the identity operator). We can describe the time evolution induced by the Kraus operator $\mathcal{V}_m$ on the principal system by a unitary operator $U$ acting upon the composite system $S+E$. Let $\ket{e_k}$ be an orthonormal basis for $E$. We can define the unitary operator $U$ such that \cite{Nielsen:2011:QuantumInfoBook}
\begin{equation}
   \ket{\psi^\prime} =  U\ket{\psi_S}\otimes\ket{e_0} = \sum_m \mathcal{V}_m \ket{\psi_S}\otimes\ket{e_m},
   \label{eq:unitary_evolution}
\end{equation}where $\ket{e_0}$ is some standard state of the environment and $\ket{\psi_S}$ is the initial state of the principal system. When applying the measurement $\ket{e_m}$ to the environment, the principal system becomes $\ket{\psi_S^\prime} \propto \mathcal{V}_m \ket{\psi_S}$. Therefore, the operator $\mathcal{V}_m$ is associated with the output $m$ and constitutes a measurement operator. We sequentially repeat this procedure to describe the continuous measurement \cite{Gammelmark:2014:QCRB}. We consider a continuous measurement during a time interval $[0,T]$. We discretize time by dividing the interval $[0,T]$ into $N$ equipartitioned intervals, where the time resolution is $\Delta t \equiv T / N$. At each time interval, we consider Eq.~\eqref{eq:unitary_evolution}. Then the state of the composite system at time $t=T$ is \cite{Gammelmark:2014:QCRB}
\begin{equation}
    \ket{\psi(T)}=\sum_{\boldsymbol{m}}\mathcal{V}_{m_{N-1}}...\mathcal{V}_{m_{0}}\ket{\psi_{S}}\otimes\ket{e_{m_{N-1}},...,e_{m_{0}}},
    \label{eq:psi_at_T}
\end{equation}where $\boldsymbol{m} \equiv [m_0,...,m_{N-1}]$. In Eq.~\eqref{eq:psi_at_T}, we assume that $\mathcal{V}_m$ is time-independent, leading to Markovian dynamics. We hereafter consider the limit of $N \to \infty$, where $\boldsymbol{m}$ becomes a record of the continuous measurement. For instance, in the case of a jump measurement, $m_i$ corresponds to either ``detection'' or ``no detection'' of a jump within $\Delta t$.
Depending upon $\boldsymbol{m}$, the state of the principal system $\ket{\psi_S(T)} \propto \mathcal{V}_{m_{N-1}}...\mathcal{V}_{m_{0}}\ket{\psi_S}$ is determined and is referred to as a quantum trajectory. 
For example, in Fig.~\ref{fig:method_ponch}, we show quantum trajectories and their corresponding measurement records for the jump [Fig.~\ref{fig:method_ponch}(a)] and diffusion [Fig.~\ref{fig:method_ponch}(b)] measurements.

The time evolution of the density operator $\rho$ is $\dot{\rho}=\left[\sum_{m}\mathcal{V}_{m}\rho\mathcal{V}_{m}^{\dagger}-\rho\right]/dt$, which obeys the Lindblad equation:
\begin{align}
\dot{\rho} & =
\mathcal{L}(\rho) \equiv
-i\left[H,\rho\right]+\sum_{c}\mathcal{D}\left(\rho,L_{c}\right),\label{eq:Lindblad_def}
\end{align}where $\mathcal{L}$ is the Lindblad operator, $\left[\cdot,\cdot\right]$ is the commutator, $H$ is a Hamiltonian, $\mathcal{D}\left(\rho,L\right)\equiv L\rho L^{\dagger}-\{L^{\dagger}L,\rho\}/2$ is the dissipator with $\{\cdot,\cdot\}$ being the anti-commutator, and $L_{c}$ is a jump operator. Although the Kraus operator $\mathcal{V}_m$ depends on measurements, the Lindblad equation does not depend on the continuous measurements performed. In Eq.~\eqref{eq:Lindblad_def}, the first and the second terms are referred to as coherent dynamics and dissipation, respectively. We assume that the Hamiltonian $H$ and the jump operators $L_c$ are parameterized by $\theta\in\mathbb{R}$; we express these expressions by $H_\theta$ and $L_{c,\theta}$, respectively. We define $\mathcal{L}_\theta$, which is the Lindblad operator consisting of $H_\theta$ and $L_{c,\theta}$. We consider the estimation of the parameter $\theta$ from the continuous measurement. Let $\Theta$ be an observable and $\mathbb{E}_\theta[\Theta]$ be the expectation of $\Theta$ with a parameter $\theta$. For an arbitrary positive-operator valued measure (POVM), according to the quantum Cram\'er--Rao inequality, Ref.~\cite{Hotta:2004:QEstimation} proved the following inequality: $\mathrm{Var}_{\theta}\left[\Theta\right]/\left(\partial_{\theta}\mathbb{E}_{\theta}\left[\Theta\right]\right)^{2}\ge1/\mathcal{I}_{Q}(\theta)$, where $\mathrm{Var}_{\theta}[\Theta]$ is the variance of $\Theta$ and $\mathcal{I}_{Q}(\theta)$ is the quantum Fisher information (see \cite{Paris:2009:QFI,Liu:2019:QFisherReview} for its review). This expression is a generalization of the conventional quantum Cram\'er--Rao inequality \cite{Helstrom:1976:QuantumEst}. Let $\mathcal{I}_C(\theta; \mathcal{M}_m)$ be the classical Fisher information obtained through POVM elements ${\mathcal{M}_m}$; then $\mathcal{I}_Q(\theta) = \max_{\mathcal{M}_m} \mathcal{I}_C(\theta; \mathcal{M}_m)$, indicating that the quantum Cram\'er--Rao inequality is satisfied by any quantum measurements \cite{Paris:2009:QFI,Liu:2019:QFisherReview}. 

Recently, Ref.~\cite{Gammelmark:2014:QCRB} obtained the quantum Fisher information for continuous measurements. 
For $T \to \infty$, Ref.~\cite{Gammelmark:2014:QCRB} showed that $\mathcal{I}_{Q}(\theta)=4T\left.\partial_{\theta_{1}}\partial_{\theta_{2}}\widetilde{\lambda}_{\theta_{1},\theta_{2}}\right|_{\theta_{1}=\theta_{2}=\theta}$, where $\widetilde{\lambda}_{\theta_1,\theta_2}$ is a dominant eigenvalue for $T\to\infty$ of a modified Lindblad operator $\widetilde{\mathcal{L}}_{\theta_{1},\theta_{2}}\rho \equiv -iH_{\theta_{1}}\rho+i\rho H_{\theta_{2}}+\sum_{c}L_{c,\theta_{1}}\rho L_{c,\theta_{2}}^{\dagger}-\frac{1}{2}\sum_{c}\left[L_{c,\theta_{1}}^{\dagger}L_{c,\theta_{1}}\rho+\rho L_{c,\theta_{2}}^{\dagger}L_{c,\theta_{2}}\right]$ (see Refs.~\cite{Gammelmark:2014:QCRB,Molmer:2015:HypoTest} for derivation).  
For $\theta_1 \to \theta$ and $\theta_2 \to \theta$, $\widetilde{\mathcal{L}}_{\theta_{1},\theta_{2}} \to \mathcal{L}_{\theta}$ and $\widetilde{\lambda}_{\theta_1,\theta_2} \to 0$.

\emph{QTUR of dynamical activity.}---We now derive a QTUR using the quantum Cram\'er--Rao inequality. We hereafter assume that the density matrix of the system is in a single steady state $\rho^\mathrm{ss}$ and only consider the limit of $T\to\infty$. In Ref.~\cite{Hasegawa:2019:CRI}, a TUR was derived via the classical Cram\'er--Rao inequality by considering a virtual perturbation \cite{Dechant:2018:FRI}, which affects only the time-scale of the dynamics while keeping the steady-state distribution unchanged. 
Analogously, we consider the following modified Hamiltonian and jump operator in Eq.~\eqref{eq:Lindblad_def}:
\begin{equation}
H_{\theta}=(1+\theta)H,\;\;\;\;L_{c,\theta}=\sqrt{1+\theta}L_{c}.\label{eq:H_L_scaling}
\end{equation}Since the Lindblad operator corresponding to Eq.~\eqref{eq:H_L_scaling} is given by $\mathcal{L}_{\theta}=(1+\theta)\mathcal{L}_{\theta=0}$, the dynamics of $\mathcal{L}_{\theta}$ are identical to the unmodified dynamics (i.e., the dynamics of $\theta=0$), except for the time scale. Let us consider a time-integrated observable $\Theta(\boldsymbol{m})$ satisfying
\begin{equation}
\mathbb{E}_{\theta}\left[\Theta(\boldsymbol{m}) \right]=h(\theta)\mathbb{E}_{\theta=0}\left[\Theta(\boldsymbol{m})\right],
\label{eq:obs_scaling}
\end{equation}where $h(\theta)$ is a scaling function independent of $\boldsymbol{m}$ [$h(0)=1$ should be satisfied]. Typically, it is given by $h(\theta)=1+\theta$. $\Theta(\boldsymbol{m})$ may be an arbitrary function of $\boldsymbol{m}$ as long as Eq.~\eqref{eq:obs_scaling} is satisfied. For instance, suppose an estimator counts the number of photons emitted during $[0,T]$; because the system is assumed to be in a steady state, the average number of photons emitted for $\mathcal{L}_{\theta}$ is $1+\theta$-times larger than that of $\mathcal{L}_{\theta=0}$, and hence this observable satisfies Eq.~\eqref{eq:obs_scaling} with $h(\theta)=1+\theta$. Combining the quantum Cram\'er--Rao inequality and Eq.~\eqref{eq:obs_scaling}, we find $\mathrm{Var}\left[\Theta\right]/\mathbb{E}\left[\Theta\right]^{2}\ge h^{\prime}(0)^{2}/\mathcal{I}_{Q}(0)$.
$\mathcal{I}_Q(\theta)$ can be calculated by differentiation of a dominant eigenvalue of $\widetilde{\mathcal{L}}_{\theta_1,\theta_2}$. Using eigenvalue differentiation \cite{Magnus:1985:EvDiff,Gammelmark:2014:QCRB}, we obtain
\begin{equation}
\frac{\mathrm{Var}\left[\Theta\right]}{\mathbb{E}\left[\Theta\right]^{2}}\ge\frac{h^{\prime}(0)^{2}}{T\left(\Upsilon+\Psi\right)}.\label{eq:QTUR1}
\end{equation}Here,
\begin{align}
\Upsilon&\equiv\sum_{c}\mathrm{Tr}\left[L_{c}\rho^{\mathrm{ss}}L_{c}^{\dagger}\right],\label{eq:Upsilon_def}\\
\Psi&\equiv-4\mathrm{Tr}\left[\mathcal{K}_{1}\circ\mathcal{L}_{\mathbb{P}}^{+}\circ\mathcal{K}_{2}(\rho^{\mathrm{ss}})+\mathcal{K}_{2}\circ\mathcal{L}_{\mathbb{P}}^{+}\circ\mathcal{K}_{1}(\rho^{\mathrm{ss}})\right],\label{eq:Psi_def}
\end{align}where $\mathcal{K}_{1}\equiv-iH\rho+\frac{1}{2}\sum_{c}\left(L_{c}\rho L_{c}^{\dagger}-L_{c}^{\dagger}L_{c}\rho\right)$ and $\mathcal{K}_{2}\equiv i\rho H+\frac{1}{2}\sum_{c}\left(L_{c}\rho L_{c}^{\dagger}-\rho L_{c}^{\dagger}L_{c}\right)$, and $\mathcal{L}_{\mathbb{P}}^{+}$ is 
a subspace of $\mathcal{L}^+$ that is complementary to the steady-state subspace, with $\mathcal{L}^+$ being the Moore–Penrose pseudo inverse of $\mathcal{L}$ (see \cite{Supp:PhysRev} for an explicit expression). Equation~\eqref{eq:QTUR1} is the first result of this Letter, which holds for arbitrary continuous measurements satisfying Eq.~\eqref{eq:obs_scaling} in Markovian open quantum systems. 

For simplicity, let us consider the following case:
\begin{equation}
L_{ji}=\sqrt{\eta_{ji}}\ket{\epsilon_{j}}\bra{\epsilon_{i}},\;\;\rho_{ij}^{\mathrm{ss}}=0\;\;(i\ne j),
\label{eq:Lij_def}
\end{equation}where $\ket{\epsilon_{i}}$ is the eigenbasis of the Hamiltonian $H$, $\eta_{ji}$ is a transition rate from $\ket{\epsilon_{i}}$ to $\ket{\epsilon_{j}}$ (we redefined the subscript of the jump operator from $L_c$ to $L_{ji}$), and $\rho^\mathrm{ss}_{ij} \equiv \braket{\epsilon_i|\rho^\mathrm{ss}|\epsilon_j}$. The off-diagonal elements of the steady-state density matrix in the energy eigenbasis are zero. These assumptions are often satisfied for quantum thermal machines \cite{Mitchison:2019:QMachine}. We obtain $\Upsilon=\sum_{i \ne j}\rho_{ii}^{\mathrm{ss}}\eta_{ji}$, corresponding to the dynamical activity in a classical Markov process, implying that $\Upsilon$ is a quantum analogue of the dynamical activity \cite{Supp:PhysRev}. Moreover, we can obtain a simpler lower bound by scaling the jump operator alone \cite{Supp:PhysRev}. In this case, $\Psi$ in Eq.~\eqref{eq:QTUR1} becomes $0$, which re-derives classical TUR. This shows that $\Psi$ quantifies the degree of the coherent dynamics in the Lindblad equation, which is also shown in a two-level atom. Therefore, Eq.~\eqref{eq:QTUR1} is a quantum generalization of a TUR \cite{Garrahan:2017:TUR,Terlizzi:2019:KUR}, which is bounded by dynamical activity. 
In classical Markov processes, a TUR bounded by the dynamical activity was derived only for discrete space systems because the dynamical activity is not well defined for continuous space. Contrastingly, Eq.~\eqref{eq:QTUR1} holds for both discrete jump and continuous diffusion cases. Recently, Ref.~\cite{Carollo:2019:QuantumLDP} proved a similar bound for quantum jump processes. The bound of Ref.~\cite{Carollo:2019:QuantumLDP} was derived for  given quantum trajectories. Therefore, their bound is obtained for a specified continuous measurement. Reference~\cite{Guarnieri:2019:QTURPRR} derived a TUR in a quantum nonequilibrium steady state using the \emph{classical} Cram{\'e}r--Rao inequality; since their TUR bounds the fluctuation of instantaneous currents (i.e., current-measurement operators), measurement effects are not explicitly incorporated.

As an example of QTUR, we consider a two-level atom driven by a classical laser field. Let $\ket{\epsilon_g}$ and $\ket{\epsilon_e}$ denote the ground and excited states, respectively. A Hamiltonian is given by  $H=\Delta\ket{\epsilon_{e}}\bra{\epsilon_{e}}+\frac{\Omega}{2}(\ket{\epsilon_{e}}\bra{\epsilon_{g}}+\ket{\epsilon_{g}}\bra{\epsilon_{e}})$,
where 
$\Delta$ is a detuning between the laser-field and the atomic-transition frequencies, and $\Omega$ is the Rabi-oscillation frequency. A jump operator is $L=\sqrt{\kappa}\ket{\epsilon_{g}}\bra{\epsilon_{e}}$, where $\kappa$ is the decay rate, and it induces a jump from $\ket{\epsilon_e}$ to $\ket{\epsilon_g}$.
We obtain the dynamical activity $\Upsilon=\kappa \rho^\mathrm{ss}_{ee}=\kappa\Omega^{2}/(4\Delta^{2}+\kappa^{2}+2\Omega^{2})$ and the coherent-dynamics contribution
\begin{equation}
\Psi=\frac{8\Omega^{4}\left[4\Delta^{4}+\Delta^{2}\left(\kappa^{2}+8\Omega^{2}\right)+\left(\kappa^{2}+2\Omega^{2}\right)^{2}\right]}{\kappa\left(4\Delta^{2}+\kappa^{2}+2\Omega^{2}\right)^{3}}.\label{eq:Psi_two_level}
\end{equation}
We first consider a jump measurement (photon detection). The quantum trajectory is given by the stochastic Schr{\"o}dinger equation (where the corresponding $\mathcal{V}_m$ is shown in \cite{Supp:PhysRev}):
\begin{align}
    d\rho&=\left(-i[H,\rho]-\frac{1}{2}\{L^{\dagger}L,\rho\}+\rho\mathrm{Tr}[L\rho L^{\dagger}]\right)dt\nonumber\\
    &+\left(\frac{L\rho L^{\dagger}}{\mathrm{Tr}[L\rho L^{\dagger}]}-\rho\right)d\mathcal{N},\label{eq:drho_jump}
\end{align}where $d\mathcal{N}$ is a noise increment and $d\mathcal{N}=1$ when a photon is detected between $t$ and $t+dt$ and $d\mathcal{N}=0$ otherwise. 
$\boldsymbol{m} = [m_{0},...,m_{N-1}]$ in Eq.~\eqref{eq:psi_at_T} corresponds to $[\Delta \mathcal{N}_0,...,\Delta \mathcal{N}_{N-1}]$. The average of this quantity reads $\mathbb{E}[d\mathcal{N}] = \mathrm{Tr}[L\rho^\mathrm{ss} L^\dagger]dt$. We consider an observable $\Theta_\mathcal{N} \equiv \int_0^T d\mathcal{N}$, which counts the number of photons emitted within the interval $[0,T]$. Since $\mathbb{E}_\theta[d\mathcal{N}] = (1+\theta)\mathbb{E}_{\theta=0}[d\mathcal{N}]$ and thus $\mathbb{E}_\theta[\Theta_\mathcal{N}] = (1+\theta)\mathbb{E}_{\theta=0}[\Theta_\mathcal{N}]$, $\Theta_\mathcal{N}$ satisfies the QTUR of Eq.~\eqref{eq:QTUR1} with $h^\prime(0) = 1$. 

We next consider a diffusion measurement (homodyne detection). A quantum trajectory of the diffusion measurement is given by a quantum-state diffusion (the corresponding $\mathcal{V}_m$ is shown in \cite{Supp:PhysRev}):
\begin{align}
    d\rho&=\left(-i[H,\rho]-\frac{1}{2}\{L^{\dagger}L,\rho\}+L\rho L^{\dagger}\right)dt\nonumber\\
    &+\left(L\rho+\rho L^{\dagger}-\mathrm{Tr}\left[L\rho+\rho L^{\dagger}\right]\rho\right)dW,\label{eq:drho_diffusion}
\end{align}where $W$ is the standard Wiener process. The measurement result is given by \cite{Gammelmark:2013:Bayes} $d\mathcal{Y}=\mathrm{Tr}[L\rho+\rho L^{\dagger}]dt+dW$. $\boldsymbol{m} = [m_{0},...,m_{N-1}]$ in Eq.~\eqref{eq:psi_at_T} corresponds to $[\Delta \mathcal{Y}_0,...,\Delta \mathcal{Y}_{N-1}]$. We consider an observable $\Theta_\mathcal{Y}\equiv \int_0^T d\mathcal{Y}$. Since $\mathbb{E}_{\theta}[\Theta_\mathcal{Y}]=\int_{0}^{T}\mathrm{Tr}[L_{\theta}\rho^{\mathrm{ss}}+\rho^{\mathrm{ss}}L_{\theta}^{\dagger}]dt=\sqrt{1+\theta}\mathbb{E}_{\theta=0}[\Theta_\mathcal{Y}]$ ($h(\theta)=\sqrt{1+\theta}$ in Eq.~\eqref{eq:obs_scaling}), $\Theta_\mathcal{Y}$ satisfies the QTUR of Eq.~\eqref{eq:QTUR1} with $h^{\prime}(0)=1/2$. Therefore, the lower bound of the diffusion measurement is $1/4$-times smaller than that of the jump measurement. 

\begin{figure}
\includegraphics[width=9cm]{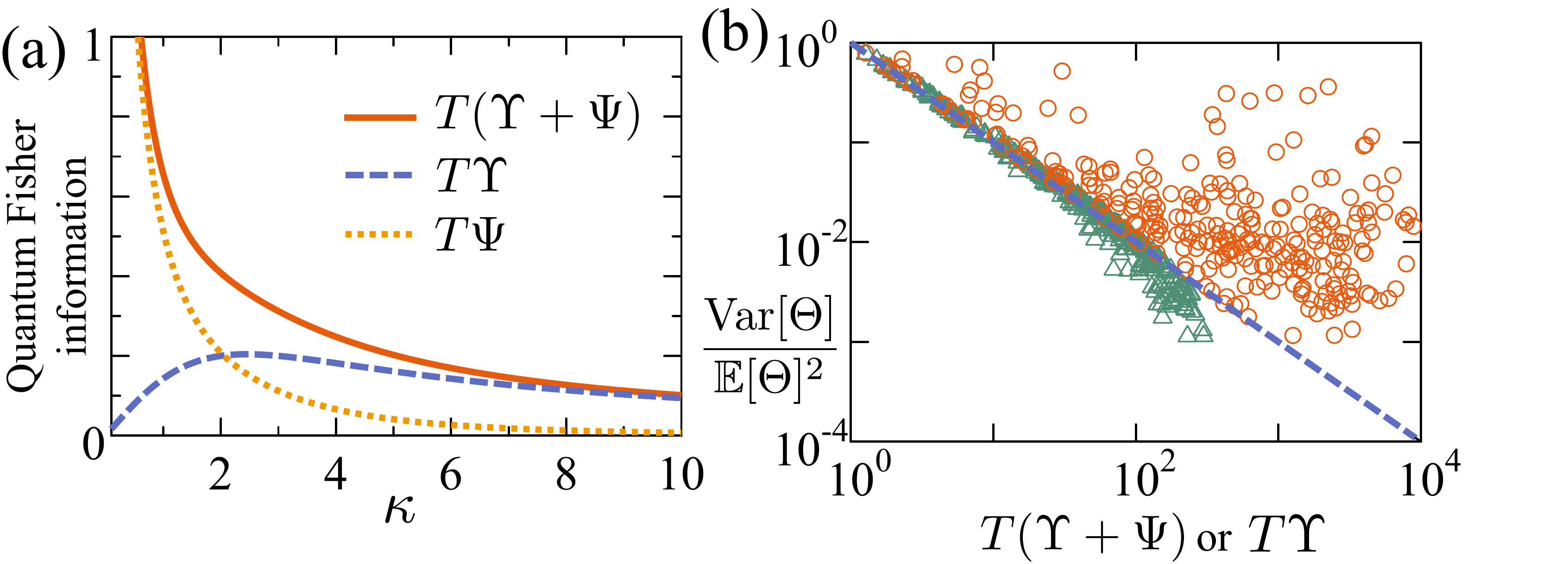}

\caption{
Quantum Fisher information and the results of computer simulations of jump measurement. 
(a) The quantum Fisher information $\mathcal{I}_Q(0) = T(\Upsilon + \Psi)$ (solid line), $T\Upsilon$ (dashed line),
and $T\Psi$ (dotted line) as a function of $\kappa$, where $T=1$, $\Omega = 1$, and $\Delta = 1$.
(b) $\mathrm{Var}[\Theta_\mathcal{N}]/\mathbb{E}[\Theta_\mathcal{N}]$ 
as a function of $T(\Upsilon + \Psi)$ (circles) and $T\Upsilon$ (triangles) for the jump measurement,
where $\Delta \in [0.1,10.0]$, $\Omega \in [0.1,10.0]$, $\kappa \in [0.1,10.0]$, and $T=1000$.
The dashed line corresponds to $1/[T(\Upsilon + \Psi)]$ for the circles
and $1/[T\Upsilon]$ for the triangles. 
\label{fig:two_atom}}
\end{figure}

We verify the QTUR of Eq.~\eqref{eq:QTUR1} for the two-level atom with a computer simulation \cite{Molmer:1993:MonteCarlo,Daley:2014:QJReview}. We first plot $\mathcal{I}_Q(0) = T(\Upsilon + \Psi)$ (solid line), $T\Upsilon$ (dashed line), and $T\Psi$ (dotted line) as a function of $\kappa$ in Fig.~\ref{fig:two_atom}(a) (parameters are shown in the caption of Fig.~\ref{fig:two_atom}(a)). From Fig.~\ref{fig:two_atom}(a), when $\kappa$ becomes larger (i.e., more frequent jumps), the dynamical activity $\Upsilon$ is dominant in the quantum Fisher information $\mathcal{I}_Q(0)$. For $\kappa \to 0$, $\Upsilon \to 0$ and the coherent dynamics contribution $\Psi$ becomes the major portion of $\mathcal{I}_Q(0)$. We numerically check the QTUR for the jump measurement by randomly generating $\kappa$, $\Omega$, and $\Delta$ (the ranges of the parameters are shown in the caption of Fig.~\ref{fig:two_atom}(b)) and calculate $\mathrm{Var}[\Theta_\mathcal{N}]/\mathbb{E}[\Theta_\mathcal{N}]^2$. In Fig.~\ref{fig:two_atom}(b), the circles denote $\mathrm{Var}[\Theta_\mathcal{N}]/\mathbb{E}[\Theta_\mathcal{N}]^2$ as a function of $\mathcal{I}_Q(0)$ and the lower bound of Eq.~\eqref{eq:QTUR1} is shown by the dashed line. We confirm that all realizations satisfy the QTUR, which verifies Eq.~\eqref{eq:QTUR1}. In a classical case \cite{Garrahan:2017:TUR,Terlizzi:2019:KUR}, the lower bound arises from the dynamical activity alone (i.e., $T\Upsilon$). Thus, we also check whether $\mathrm{Var}[\Theta_\mathcal{N}]/\mathbb{E}[\Theta_\mathcal{N}]^2$ can be bounded only by $T\Upsilon$. In Fig.~\ref{fig:two_atom}(b), the triangles denote $\mathrm{Var}[\Theta_\mathcal{N}]/\mathbb{E}[\Theta_\mathcal{N}]^2$ as a function of $T\Upsilon$, where the dashed line describes $1/(T\Upsilon)$. Clearly, some realizations are below $1/(T\Upsilon)$, indicating that the lower bound of the QTUR is below the classical bound \cite{Garrahan:2017:TUR,Terlizzi:2019:KUR}. Similar enhancement of precision has been reported for quantum jump processes \cite{Carollo:2019:QuantumLDP}, and for classical systems with periodic driving \cite{Barato:2018:PeriodicTUR} or magnetic fields \cite{Macieszczak:2018:TURLR}. We also performed a computer simulation for the diffusion measurement and verified the bound (see \cite{Supp:PhysRev}).

\begin{figure}
\includegraphics[width=8cm]{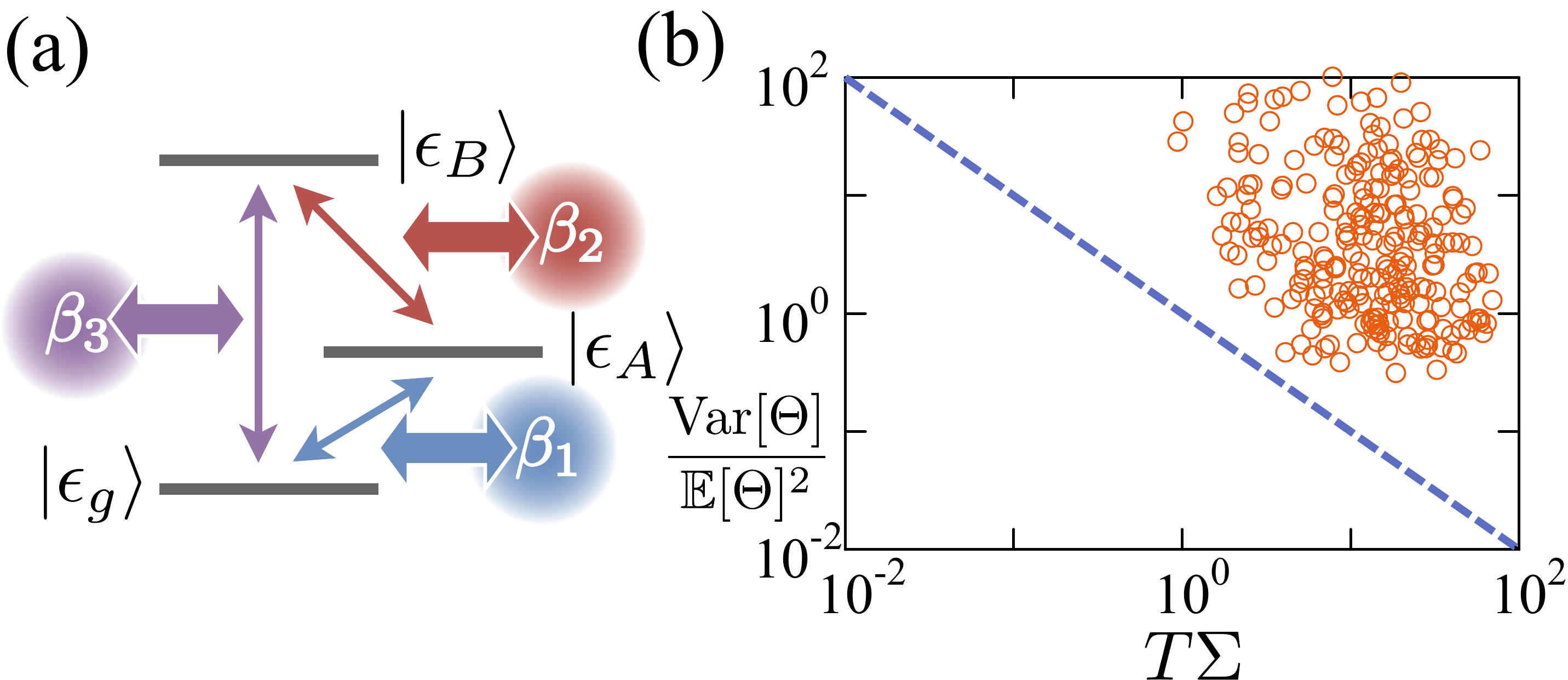}
\caption{
Illustration of the model and results of computer simulation for the thermal machine.
(a) Thermal machine consisting of three levels $\ket{\epsilon_A}$, $\ket{\epsilon_B}$, and $\ket{\epsilon_g}$.
The transitions between each of the states are coupled with heat reservoirs with the inverse temperature $\beta_r$ ($r=1,2,3$). 
(b) $\mathrm{Var}[\Theta_C^\prime]/\mathbb{E}[\Theta_C^\prime]$ (circles)
as a function of $T\Sigma$ for the transformed jump measurement,
where 
$\beta_1\in[0.1,1.0]$,$\beta_2\in[0.01,0.1]$,$\beta_3\in[1.0,10.0]$,
$\omega_1\in[1.0,10.0]$,$\omega_2\in[1.0,10.0]$,
$R^\prime_{ij}\in[0.0,1.0]$, $|\zeta_{ij}|\in[0.0,0.2]$,
$\gamma=0.1$, and $T=100$.
$R^\prime_{ji}$ and $\zeta_{ji}$ are defined for $i\ne j$ and satisfy
$R^\prime_{ji} = -R^\prime_{ij}$ and $|\zeta_{ji}| = |\zeta_{ij}|$. 
The dashed line corresponds to the lower bound $2/[T\Sigma]$. 
\label{fig:thermal_machine}}
\end{figure}

\emph{QTUR of entropy production.}---
Employing a scaling different from Eq.~\eqref{eq:H_L_scaling}, we can bound $\mathrm{Var}[\Theta]/\mathbb{E}[\Theta]^2$ by the entropy production. Again, we assume that the system satisfies the conditions of Eq.~\eqref{eq:Lij_def}. Moreover, we assume that whenever $\eta_{ji} >0$, $\eta_{ij}>0$ should be satisfied. Inspired by Ref.~\cite{Dechant:2019:MTUR}, we consider the following modified process instead of Eq.~\eqref{eq:H_L_scaling}:
\begin{align}
L_{ji,\theta}&=\sqrt{\eta_{ji}\left[1+\theta\left(1-\sqrt{\frac{\eta_{ij}\rho_{jj}^{\mathrm{ss}}}{\eta_{ji}\rho_{ii}^{\mathrm{ss}}}}\right)\right]}\ket{\epsilon_{j}}\bra{\epsilon_{i}}\; (i \ne j).\label{eq:L_scaling_EP}
\end{align}
With Eq.~\eqref{eq:L_scaling_EP}, the steady-state density remains unchanged. Repeating a similar calculation to the dynamical-activity case (see \cite{Supp:PhysRev} for details), an observable $\Theta$ satisfying Eq.~\eqref{eq:obs_scaling} obeys
\begin{equation}
\frac{\mathrm{Var}\left[\Theta\right]}{\mathbb{E}\left[\Theta\right]^{2}}\ge\frac{2h^{\prime}(0)^{2}}{T\Sigma},
\label{eq:QTUR2}
\end{equation}where $\Sigma \equiv \sum_{i\ne j}\rho_{ii}^{\mathrm{ss}}\eta_{ji}\ln\left[\rho_{ii}^{\mathrm{ss}}\eta_{ji}/\left(\rho_{jj}^{\mathrm{ss}}\eta_{ij}\right)\right]$. Equation~\eqref{eq:QTUR2} is the second result of this Letter. The expression of $\Sigma$ is identical to the entropy production in stochastic thermodynamics \cite{Seifert:2012:FTReview}; 
therefore, our approach re-derives the classical TUR \cite{Barato:2015:UncRel,Gingrich:2016:TUP} but its applicability is broader than that of classical counterpart, as detailed below. 

As an example, we consider a quantum thermal machine. Such machines are the basis for quantum clocks and thus it is important to consider their precision \cite{Erker:2017:QClockTUR,Mitchison:2019:QMachine}. 
Specifically, we employ a three-level thermal machine powered by three heat reservoirs at different inverse temperatures $\beta_r$ ($r=1,2,3$) \cite{Mitchison:2019:QMachine,Paule:2018:PhDBook}. Each transition is coupled with each of the heat reservoirs (Fig.~\ref{fig:thermal_machine}(a)). The Hamiltonian is $H = \omega_3 \ket{\epsilon_B}\bra{\epsilon_B} + \omega_1\ket{\epsilon_A}\bra{\epsilon_A}$, where $\omega_1$, $\omega_2$, and $\omega_3 = \omega_1 + \omega_2$ are energy gaps between $\ket{\epsilon_A}\leftrightarrow \ket{\epsilon_g}$, $\ket{\epsilon_B}\leftrightarrow \ket{\epsilon_A}$, and $\ket{\epsilon_B}\leftrightarrow \ket{\epsilon_g}$, respectively. 
Let $\dot{Q}_r$ be the heat current from the $r$th reservoir with temperature $\beta_r$. We assume that the dynamics of the density operator $\rho$ obey the Lindblad equation $\dot{\rho}=-i[H,\rho]+\sum_{i\ne j}\mathcal{D}(\rho,L_{ji})$, where $L_{ji}$ is defined in Eq.~\eqref{eq:Lij_def} with $\eta_{gA}=\gamma(n^\mathrm{th}_1 + 1)$, $\eta_{Ag}=\gamma n^\mathrm{th}_1$, $\eta_{AB}=\gamma (n^\mathrm{th}_2 + 1)$, $\eta_{BA} = \gamma n^\mathrm{th}_2$, $\eta_{gB}=\gamma (n^\mathrm{th}_3 + 1)$, and $\eta_{Bg}=\gamma n^\mathrm{th}_3$ [$n^\mathrm{th}_r\equiv (e^{\beta_r\omega_r} - 1)^{-1}$ and $\gamma$ is the decay rate].
The entropy production is $\Phi= -\sum_{r=1}^{3} \beta_i \dot{Q}_r$, where $\dot{Q}_r$ is the heat flux entering from the $r$th reservoir \cite{Alicki:1979:HeatEngine,Boukobza:2006:Thermo}, and satisfies $\Phi = \Sigma$ \cite{Supp:PhysRev}. 
Therefore, the classical entropy production $\Sigma$ corresponds to the entropy production in the quantum heat engine $\Phi$. 

We first consider a standard jump measurement. The quantum trajectory is given by a stochastic Schr{\"o}dinger equation:
\begin{align}
    d\rho&=-i[H,\rho]dt+\sum_{i\ne j}\left(\rho\mathrm{Tr}\left[L_{ji}\rho L_{ji}^{\dagger}\right]-\frac{\left\{ L_{ji}^{\dagger}L_{ji},\rho\right\} }{2}\right)dt\nonumber\\&+\sum_{i\ne j}\left(\frac{L_{ji}\rho L_{ji}^{\dagger}}{\mathrm{Tr}[L_{ji}\rho L_{ji}^{\dagger}]}-\rho\right)d\mathcal{N}_{ji}.
    \label{eq:SSE_refrige}
\end{align}We consider the observable $\Theta_C \equiv \sum_{i\ne j}  R_{ji} \int_0^T d\mathcal{N}_{ji}$, where $R_{ji} = -R_{ij}$ and $R_{ji} \in \mathbb{R}$. $\Theta_C$ satisfies the scaling condition of Eq.~\eqref{eq:obs_scaling} and thus the QTUR of Eq.~\eqref{eq:QTUR2}. Because the dynamics of Eq.~\eqref{eq:SSE_refrige} are jumps between energy eigenstates that are equivalent to classical dynamics, $\Theta_C$ trivially satisfies Eq.~\eqref{eq:QTUR2}. 

We next consider a transformed jump measurement \cite{Wiseman:1999:QJump}. The Lindblad equation is invariant under the transformation $L_{ji}^{\prime}=L_{ji}+\zeta_{ji}\mathbb{I}$ and $H^{\prime}=H-\frac{i}{2}\sum_{i\ne j}\left[\zeta_{ji}^{*}L_{ji}-\zeta_{ji}L_{ji}^{\dagger}\right]$, where $\zeta_{ji} \in \mathbb{C}$ is a parameter. $\zeta_{ji}=0$ for all $i$ and $j$  recovers the standard jump measurement. 
Thus, we can consider a transformed stochastic Schr{\"o}dinger equation, where $H$ and $L_{ji}$ are replaced with $H^\prime$ and $L_{ji}^\prime$, respectively, in Eq.~\eqref{eq:SSE_refrige}, and we define $d\mathcal{N}_{ji}^\prime$ as a noise increment in the transformed equation. Quantum trajectories are no longer simple jump processes between energy eigenstates \cite{Supp:PhysRev}. We consider an observable $\Theta^\prime_C \equiv \sum_{i\ne j}  R^\prime_{ji} \int_0^T d\mathcal{N}_{ji}^\prime$, where $R^\prime_{ji} = -R^\prime_{ij}$, for the transformed equation. When $|\zeta_{ji}| = |\zeta_{ij}|$ for all $i$ and $j$, $\Theta^\prime_C$ satisfies the QTUR of Eq.~\eqref{eq:QTUR2} \cite{Supp:PhysRev}.

We verify the QTUR of Eq.~\eqref{eq:QTUR2} for the transformed jump measurement (i.e., $\zeta_{ji}\ne 0$ and 
$|\zeta_{ij}| = |\zeta_{ji}|$) via a computer simulation and numerically check the QTUR by randomly generating $\beta_r$, $\omega_r$, $R^\prime_{ji}$, and $\zeta_{ji}$ (parameters are shown in the caption of Fig.~\ref{fig:thermal_machine}(b)) and calculating $\mathrm{Var}[\Theta_C^\prime]/\mathbb{E}[\Theta_C^\prime]^2$. In Fig.~\ref{fig:thermal_machine}(b), the circles denote $\mathrm{Var}[\Theta_C^\prime]/\mathbb{E}[\Theta_C^\prime]^2$ as a function of the entropy production $T\Sigma$ and the lower bound of Eq.~\eqref{eq:QTUR2} is shown by a dashed line. We confirm that all realizations satisfy the QTUR, verifying Eq.~\eqref{eq:QTUR2}. Although the bound of Eq.~\eqref{eq:QTUR2} itself is identical to the classical TUR \cite{Barato:2015:UncRel,Gingrich:2016:TUP}, our QTUR provides the lower bound for arbitrary measurements with the scaling condition. No matter how we measure the thermal machine, an observable satisfying the scaling relation [Eq.~\eqref{eq:obs_scaling}] should obey the QTUR of Eq.~\eqref{eq:QTUR2}, which cannot be deduced from classical TURs. 
We also note observables not satisfying the scaling condition of Eq.~\eqref{eq:obs_scaling}. As demonstrated in the example, although the scaling condition is satisfied for typical measurement schemes, such as jump and diffusion measurements, this is not the case for general continuous measurements. For such cases, inequalities of Eq.~\eqref{eq:QTUR1} and \eqref{eq:QTUR2} hold with $\mathbb{E}[\Theta]$ replaced by $\partial_\theta \mathbb{E}_\theta[\Theta]$.

\emph{Conclusion.}---In this Letter, we have derived the QTUR from the quantum Cram\'er-Rao inequality. The QTUR holds for arbitrary continuous measurements satisfying the scaling condition. We expect the present study to form a basis for obtaining uncertainty relations in the quantum regime. 

\emph{Acknowledgements.}---The present study was supported by the Ministry of Education, Culture, Sports, Science and Technology (MEXT) KAKENHI Grant No.~JP19K12153.

\end{document}


\title{Supplementary Material of \\
``Quantum Thermodynamic Uncertainty Relation for Continuous Measurement''}
\author{Yoshihiko Hasegawa}
\affiliation{Department of Information and Communication Engineering, Graduate
School of Information Science and Technology, The University of Tokyo,
Tokyo 113-8656, Japan}

\maketitle
This supplementary material describes the calculations introduced in the main text. Equation and figure numbers are prefixed with S (e.g., Eq.~(S1) or Fig.~S1). Numbers without this prefix (e.g., Eq.~(1) or Fig.~1) refer to items in the main text.

\section{Basics}
\subsection{Liouville-space representation}

First, we introduce the vectorization of quantum operators following \cite{Landi:2018:TextBook}. Typically, quantum dynamics are described in a Hilbert-space representation. There, a density matrix $\rho$ is
\[
\rho=\sum_{i,j}\rho_{ij}\ket{i}\bra{j},
\]where $\ket{i}$ is some orthonormal basis in the Hilbert space. We introduce a vectorized form of $\rho$ as
\begin{equation}
\mathrm{vec}(\rho) \equiv \sum_{i,j}\rho_{ij}\ket{j}\otimes\ket{i},\label{eq:vec_def}
\end{equation}which now belongs to a Liouville space. When the dimensionality of the Hilbert space is $d$, its corresponding Liouville space has dimensionality $d^2$. Let $A$, $B$, and $C$ be matrices in the Hilbert space. With Eq.~\eqref{eq:vec_def}, the following relation holds
\begin{equation}
\mathrm{vec}(ABC)=(C^{\top}\otimes A)\mathrm{vec}(B),\label{eq:ABC_transform}
\end{equation}where $\top$ is the matrix transpose. With vectorization, the Hilbert--Schmidt inner product becomes
\[
\left\langle A,B\right\rangle \equiv\mathrm{Tr}[A^{\dagger}B]=\mathrm{vec}(A)^{\dagger}\mathrm{vec}(B)=\left\langle \mathrm{vec}(A),\mathrm{vec}(B)\right\rangle ,
\]which is the inner product between two vectors $\mathrm{vec}(A)$ and $\mathrm{vec}(B)$. Consider the following Lindblad equation:
\begin{equation}
\dot{\rho} =\mathcal{L}(\rho) \equiv -i[H,\rho]+\sum_{c}\left[L_{c}\rho L_{c}^{\dagger}-\frac{1}{2}\left\{ L_{c}^{\dagger}L_{c}\rho+\rho L_{c}^{\dagger}L_{c}\right\} \right],\label{eq:Lindblad_def}
\end{equation}where $\mathcal{L}$ is the Lindblad operator, $H$ is the Hamiltonian, and $L_c$ is the jump operator. Using Eq.~\eqref{eq:vec_def}, Eq.~\eqref{eq:Lindblad_def} is converted into
\[
\frac{d}{dt}\mathrm{vec}(\rho)=\hat{\mathcal{L}}\mathrm{vec}(\rho),
\]where $\hat{\mathcal{L}}$ is a matrix representation (Liouville-space representation) of $\mathcal{L}$ obtained through Eq.~\eqref{eq:ABC_transform}, as
\begin{equation}
\hat{\mathcal{L}} \equiv -i(\mathbb{I}\otimes H-H^{\top}\otimes\mathbb{I})+\sum_{c}\left[L_{c}^{*}\otimes L_{c}-\frac{1}{2}\mathbb{I}\otimes L_{c}^{\dagger}L_{c}-\frac{1}{2}(L_{c}^{\dagger}L_{c})^{\top}\otimes\mathbb{I}\right].
\label{eq:L_matrix_def}
\end{equation}
Here, a superscript $*$ denotes complex conjugate and $\mathbb{I}$ is the identity operator. Now, the Lindblad equation becomes a linear differential equation and we can compute the steady-state density matrix from $\hat{\mathcal{L}}$. The right and left eigenvectors corresponding to a zero eigenvalue are
\begin{align}
\hat{\mathcal{L}}\mathrm{vec}(\rho^{\mathrm{ss}}) & =0,\label{eq:right_ev_def}\\
\mathrm{vec}(\mathbb{I})^{\dagger}\hat{\mathcal{L}} & =0,\label{eq:left_ev_def}
\end{align}where $\rho^\mathrm{ss}$ is the steady-state density matrix. 

\subsection{Eigenvalue derivative}

We show an expression for the eigenvalue derivative based on Refs.~\cite{Magnus:1985:EvDiff,Gammelmark:2014:QCRB}. Let us consider an eigenvalue problem. Let $A(\bm{\theta})$ be a matrix parametrized by a vector $\bm{\theta}$. We assume that the eigenvalues are not degenerate. Right and left eigenvectors ($u(\bm{\theta})$ and $v(\bm{\theta})$, respectively) of an eigenvalue $\lambda(\bm{\theta})$ satisfy
\begin{align}
A(\bm{\theta})u(\bm{\theta}) & =\lambda(\bm{\theta})u(\bm{\theta}),\label{eq:right_eigenvector}\\
A(\bm{\theta})^{\dagger}v(\bm{\theta}) & =\lambda^{*}(\bm{\theta})v(\bm{\theta}).\label{eq:left_eigenvector}
\end{align}We wish to find the derivative of $\lambda(\bm{\theta})$ around some chosen value $\bm{\theta}_{0}$. Specifically, we focus on an eigenvalue that vanishes at $\bm{\theta}_{0}$, i.e., $\lambda(\bm{\theta}_{0})=0$. We impose the following normalization constraints:
\begin{align}
\left\langle u(\bm{\theta}_{0}),u(\bm{\theta})\right\rangle  & =1,\label{eq:u_u_normalization}\\
\left\langle u(\bm{\theta}_{0}),v(\bm{\theta}_{0})\right\rangle  & =1.\label{eq:u_v_normalization}
\end{align}Now let us introduce the following notation:
\[
A_{i}(\bm{\theta}_{0})\equiv\left.\frac{\partial}{\partial\theta_{i}}A(\bm{\theta})\right|_{\bm{\theta}=\bm{\theta}_{0}},\;\;\;A_{ij}(\bm{\theta}_{0})\equiv\left.\frac{\partial^{2}}{\partial\theta_{i}\partial\theta_{j}}A(\bm{\theta})\right|_{\bm{\theta}=\bm{\theta}_{0}}.
\]From Ref.~\cite{Magnus:1985:EvDiff,Gammelmark:2014:QCRB}, the derivative of $\lambda(\bm{\theta})$ is
\begin{align}
\left.\frac{\partial^{2}}{\partial\theta_{i}\partial\theta_{j}}\lambda(\bm{\theta})\right|_{\bm{\theta}=\bm{\theta}_{0}} & =\mathcal{X}+\mathcal{Z}_{1}+\mathcal{Z}_{2},\label{eq:ev_derivative}
\end{align}where
\begin{align}
\mathcal{X} & \equiv\left\langle v(\bm{\theta}_{0}),A_{ij}(\bm{\theta}_{0})u(\bm{\theta}_{0})\right\rangle ,\label{eq:X_def}\\
\mathcal{Z}_{1} & \equiv-\left\langle v(\bm{\theta}_{0}),A_{i}(\bm{\theta}_{0})\mathbb{P}A(\bm{\theta}_{0})^{+}\mathbb{P}A_{j}(\bm{\theta}_{0})u(\bm{\theta}_{0})\right\rangle ,\label{eq:Z1_def}\\
\mathcal{Z}_{2} & \equiv-\left\langle v(\bm{\theta}_{0}),A_{j}(\bm{\theta}_{0})\mathbb{P}A(\bm{\theta}_{0})^{+}\mathbb{P}A_{i}(\bm{\theta}_{0})u(\bm{\theta}_{0})\right\rangle .\label{eq:Z2_def}
\end{align}Here, $A^{+}$ is the Moore--Penrose pseudo inverse of $A$ and $\mathbb{P}$ is a projector defined by
\begin{equation}
\mathbb{P}\equiv\mathbb{I}-u(\bm{\theta}_{0})v(\bm{\theta}_{0})^{\dagger}.\label{eq:P_def}
\end{equation}
\section{Derivations}

\subsection{Bound by dynamical activity}

We derive the QTUR bounded by the dynamical activity. We consider the following scaling in the main text {[}Eq.~\HULUscaling{}{]}:
\begin{equation}
H_{\theta}=(1+\theta)H,\;\;\;L_{c,\theta}=\sqrt{1+\theta}L_{c}.\label{eq:H_L_scaling}
\end{equation}The corresponding modified Lindblad operator is 
\begin{align}
\widetilde{\mathcal{L}}_{\theta_{1},\theta_{2}}(\rho) & =-i\left[\left(1+\theta_{1}\right)H\rho-\left(1+\theta_{2}\right)\rho H\right]+\sqrt{\left(1+\theta_{1}\right)\left(1+\theta_{2}\right)}\sum_{c}L_{c}\rho L_{c}^{\dagger}\nonumber \\
 & -\frac{1+\theta_{1}}{2}\sum_{c}L_{c}^{\dagger}L_{c}\rho-\frac{1+\theta_{2}}{2}\sum_{c}\rho L_{c}^{\dagger}L_{c}.\label{eq:mod_Lindblad_op_DA}
\end{align}Let $\widehat{\widetilde{\mathcal{L}}}_{\theta_{1},\theta_{2}}$ be a matrix representation (Liouville-space representation) of the modified Lindblad operator, which is
obtained by Eq.~\eqref{eq:ABC_transform}. 
We wish to obtain the derivative of an eigenvalue of $\widehat{\widetilde{\mathcal{L}}}_{\theta_{1},\theta_{2}}$ through Eq.~\eqref{eq:ev_derivative}. For $A(\bm{\theta})=\widehat{\widetilde{\mathcal{L}}}_{\theta_{1},\theta_{2}}$, we find the following relation from Eqs.~\eqref{eq:right_ev_def}, \eqref{eq:left_ev_def},  \eqref{eq:u_u_normalization}, and \eqref{eq:u_v_normalization}:
\begin{align*}
u(\bm{\theta}_{0})&=k_{u}\mathrm{vec}(\rho^{\mathrm{ss}}),\\
v(\bm{\theta}_{0})&=k_{v}\mathrm{vec}(\mathbb{I}),
\end{align*}where $k_{u}\equiv 1/\sqrt{\left\langle \mathrm{vec}(\rho^{\mathrm{ss}}),\mathrm{vec}(\rho^{\mathrm{ss}})\right\rangle }$ and $k_{v}\equiv \sqrt{\left\langle \mathrm{vec}(\rho^{\mathrm{ss}}),\mathrm{vec}(\rho^{\mathrm{ss}})\right\rangle }$ are normalization constants. From Eq.~\eqref{eq:ev_derivative}, the eigenvalue differentiation is given by
\begin{align}
\left.\frac{\partial^{2}}{\partial\theta_{1}\partial\theta_{2}}\widetilde{\lambda}_{\theta_{1},\theta_{2}}\right|_{\theta_{1}=\theta_{2}=\theta} & =\mathcal{X}+\mathcal{Z}_{1}+\mathcal{Z}_{2},\label{eq:lambda_X_Z}
\end{align}where $\widetilde{\lambda}_{\theta_1,\theta_2}$ is a dominant eigenvalue of a modified Lindblad operator $\widetilde{\mathcal{L}}_{\theta_{1},\theta_{2}}$. $\widetilde{\lambda}_{\theta_1,\theta_2}$ smoothly converges to $0$ for $\theta_1\to 0$ and $\theta_2\to 0$. From Eq.~\eqref{eq:mod_Lindblad_op_DA}, we obtain
\begin{align}
\mathcal{X} & =\mathrm{Tr}\left[\frac{\partial^{2}}{\partial\theta_{1}\partial\theta_{2}}\widetilde{\mathcal{L}}_{\theta_{1},\theta_{2}}(\rho^{\mathrm{ss}})\right]_{\theta_{1}=\theta_{2}=\theta}=\frac{1}{4}\sum_{c}\mathrm{Tr}\left[L_{c}\rho^{\mathrm{ss}}L_{c}^{\dagger}\right].
\label{eq:X_DA_def}
\end{align}Since $\mathrm{Tr}[L_{c}\rho^{\mathrm{ss}}L_{c}^{\dagger}]^{*}=\mathrm{Tr}[(L_{c}\rho^{\mathrm{ss}}L_{c}^{\dagger})^{\dagger}]=\mathrm{Tr}[L_{c}\rho^{\mathrm{ss}}L_{c}^{\dagger}]$, we find that $\mathcal{X}$ is real. $\mathcal{Z}_1$ and $\mathcal{Z}_2$ are given by
\begin{align*}
\mathcal{Z}_{1} & =-\left\langle \mathrm{vec}(\mathbb{I}),\hat{\mathcal{K}}_{1}\left(\mathbb{I}-\mathrm{vec}(\rho^{\mathrm{ss}})\mathrm{vec}(\mathbb{I})^{\dagger}\right)\hat{\mathcal{L}}^{+}\left(\mathbb{I}-\mathrm{vec}(\rho^{\mathrm{ss}})\mathrm{vec}(\mathbb{I})^{\dagger}\right)\hat{\mathcal{K}}_{2}\mathrm{vec}(\rho^{\mathrm{ss}})\right\rangle ,\\
\mathcal{Z}_{2} & =-\left\langle \mathrm{vec}(\mathbb{I}),\hat{\mathcal{K}}_{2}\left(\mathbb{I}-\mathrm{vec}(\rho^{\mathrm{ss}})\mathrm{vec}(\mathbb{I})^{\dagger}\right)\hat{\mathcal{L}}^{+}\left(\mathbb{I}-\mathrm{vec}(\rho^{\mathrm{ss}})\mathrm{vec}(\mathbb{I})^{\dagger}\right)\hat{\mathcal{K}}_{1}\mathrm{vec}(\rho^{\mathrm{ss}})\right\rangle ,
\end{align*}where $\hat{\mathcal{L}}^+$ denotes the Moore-Penrose pseudo inverse of $\hat{\mathcal{L}}$, and $\hat{\mathcal{K}}_{1}$ and $\hat{\mathcal{K}}_{2}$ are matrix representations (Liouville-space representation) of
\begin{align*}
\mathcal{K}_{1} & \equiv-iH\rho+\frac{1}{2}\sum_{c}\left(L_{c}\rho L_{c}^{\dagger}-L_{c}^{\dagger}L_{c}\rho\right),\\
\mathcal{K}_{2} & \equiv i\rho H+\frac{1}{2}\sum_{c}\left(L_{c}\rho L_{c}^{\dagger}-\rho L_{c}^{\dagger}L_{c}\right).
\end{align*}As noted in the main text, for $T\to \infty$, Ref.~\cite{Gammelmark:2014:QCRB} found that the quantum Fisher information for continuous measurements is given by
\begin{equation}
\mathcal{I}_{Q}(\theta=0)=4T\left[\frac{\partial^{2}}{\partial\theta_{1}\partial\theta_{2}}\widetilde{\lambda}_{\theta_{1},\theta_{2}}\right]_{\theta_{1}=\theta_{2}=0}.
\label{eq:QFI_CM}
\end{equation}Substituting Eq.~\eqref{eq:lambda_X_Z} into \eqref{eq:QFI_CM}, we obtain
\begin{align*}
\mathcal{I}_{Q}(\theta=0)=4T\left[\mathcal{X}+\mathcal{Z}_{1}+\mathcal{Z}_{2}\right]=T\left[\Upsilon+\Psi\right],
\end{align*}yielding the first main result of the main text {[}Eq.~\QTURI{}{]}:
\begin{equation}
\frac{\mathrm{Var}\left[\Theta\right]}{\mathbb{E}\left[\Theta\right]^{2}}\ge\frac{h^{\prime}(0)^{2}}{T\left(\Upsilon+\Psi\right)}.\label{eq:QTUR1}
\end{equation}

Next, we limit our discussion to the following case [Eq.~\LijUdef{}]:
\begin{equation}
L_{ji}=\sqrt{\eta_{ji}}\ket{\epsilon_{j}}\bra{\epsilon_{i}},\;\;\;\rho_{ij}^{\mathrm{ss}}=0\;\;\;(i\ne j),
\label{eq:Lij_def}
\end{equation}where $\ket{\epsilon_{i}}$ is the eigenbasis of the Hamiltonian $H$, $\eta_{ji}$ is a transition rate from $\ket{\epsilon_{i}}$ to $\ket{\epsilon_{j}}$ (we redefined the subscript of the jump operator from $L_c$ to $L_{ji}$), and $\rho^\mathrm{ss}_{ij} \equiv \braket{\epsilon_i|\rho^\mathrm{ss}|\epsilon_j}$. The dynamics of this system are jumps between energy eigenstates. Therefore, this system corresponds to a classical Markov chain and we can derive the classical TUR through the above derivation. Now we only have to consider a scaling for the jump operator:
\begin{equation}
H_{\theta}=H,\;\;\;L_{c,\theta}=\sqrt{1+\theta}L_{c}.\label{eq:H_L_scaling_Lonly}
\end{equation}$\mathcal{X}$ in Eq.~\eqref{eq:lambda_X_Z} remains unchanged for this specific case, i.e., $\mathcal{X}$ is given by Eq.~\eqref{eq:X_DA_def}. To calculate $\mathcal{Z}_{1}$ (and $\mathcal{Z}_{2}$), we focus on $A_{j}(\bm{\theta}_{0})u(\bm{\theta}_{0})$ in Eqs.~\eqref{eq:Z1_def} and \eqref{eq:Z2_def}, which is calculated into ($j=1$)
\begin{align}
A_{1}(\bm{\theta}_{0})u(\bm{\theta}_{0})&\propto\left.\frac{\partial\widetilde{\mathcal{L}}_{\theta_{1},\theta_{2}}}{\partial\theta_{1}}\left(\rho^{\mathrm{ss}}\right)\right|_{\theta_{1}=\theta_{2}=0}\nonumber\\
&=\frac{1}{2}\sum_{i \ne j}\eta_{ji}\left(\rho_{ii}^{\mathrm{ss}}\ket{\epsilon_{j}}\bra{\epsilon_{j}}-\ket{\epsilon_{i}}\bra{\epsilon_{i}}\rho^{\mathrm{ss}}\right).
\label{eq:Au_DA}
\end{align}Since we have assumed that the density matrix does not have off-diagonal elements in the energy eigenbasis, we substitute $\rho^{\mathrm{ss}}=\sum_{i}\rho_{ii}^{\mathrm{ss}}\ket{\epsilon_{i}}\bra{\epsilon_{i}}$ to obtain
\begin{align}
A_{1}(\bm{\theta}_{0})u(\bm{\theta}_{0})&\propto \sum_{i\ne j}\eta_{ji}\left(\rho_{ii}^{\mathrm{ss}}\ket{\epsilon_{j}}\bra{\epsilon_{j}}-\rho_{ii}^{\mathrm{ss}}\ket{\epsilon_{i}}\bra{\epsilon_{i}}\right).
 \label{eq:Z_zero_DA}
\end{align}Multiplying $\bra{\epsilon_k}$ and $\ket{\epsilon_k}$ to Eq.~\eqref{eq:Z_zero_DA} from left and right, respectively, we obtain
\begin{align}
    \braket{\epsilon_{k}|\sum_{i\ne j}\eta_{ji}\left(\rho_{ii}^{\mathrm{ss}}\ket{\epsilon_{j}}\bra{\epsilon_{j}}-\rho_{ii}^{\mathrm{ss}}\ket{\epsilon_{i}}\bra{\epsilon_{i}}\right)|\epsilon_{k}}&=\sum_{i}\eta_{ki}\rho_{ii}^{\mathrm{ss}}-\sum_{j}\eta_{jk}\rho_{kk}^{\mathrm{ss}}\nonumber\\
    &=0.\label{eq:braket_zero_DA}
\end{align}The last line of Eq.~\eqref{eq:braket_zero_DA} holds because $\rho^\mathrm{ss}$ is the steady-state solution of the Lindblad equation. From Eq.~\eqref{eq:braket_zero_DA}, $\mathcal{Z}_1 = \mathcal{Z}_2 = 0$ and we obtain
\begin{equation}
\frac{\mathrm{Var}\left[\Theta\right]}{\mathbb{E}\left[\Theta\right]^{2}}\ge\frac{h^{\prime}(0)^{2}}{T\Upsilon}.
\label{eq:QTUR1_specific}
\end{equation}Equation~\eqref{eq:QTUR1_specific} re-derives the classical TUR bounded by the dynamical activity.

\subsection{Bound by entropy production}
We next derive the QTUR bounded by the entropy production. We again limit our discussion to 
Eq.~\eqref{eq:Lij_def}. Inspired by Ref.~\cite{Dechant:2019:MTUR}, we consider the following modified jump operator in the main text {[}Eq.~\LUscalingUEP{}{]}:
\begin{equation}
L_{ji,\theta}=\sqrt{\eta_{ji}\left[1+\theta\left(1-\sqrt{\frac{\eta_{ij}\rho_{jj}^{\mathrm{ss}}}{\eta_{ji}\rho_{ii}^{\mathrm{ss}}}}\right)\right]}\ket{\epsilon_{j}}\bra{\epsilon_{i}}\;\;\;\;(i \ne j).\label{eq:EP_scaling}
\end{equation}The derivative of $L_{ji,\theta}$ is
\begin{equation}
\left.\frac{\partial}{\partial\theta}L_{ji,\theta}\right|_{\theta=0}=\frac{1}{2}\sqrt{\eta_{ji}}\left(1-\sqrt{\frac{\eta_{ij}\rho_{jj}^{\mathrm{ss}}}{\eta_{ji}\rho_{ii}^{\mathrm{ss}}}}\right)\ket{\epsilon_{j}}\bra{\epsilon_{i}}.
\label{eq:Lij_derivative}
\end{equation}Similar to the dynamical-activity case considered above, the derivative of $\widetilde{\lambda}_{\theta_{1},\theta_{2}}$ is given by Eq.~\eqref{eq:lambda_X_Z}, where $\mathcal{X}$ is
\begin{align*}
\mathcal{X}&\equiv\mathrm{Tr}\left[\frac{\partial^{2}}{\partial\theta_{1}\partial\theta_{2}}\widetilde{\mathcal{L}}_{\theta_{1},\theta_{2}}(\rho^{\mathrm{ss}})\right]_{\theta_{1}=\theta_{2}=0}=\sum_{i\ne j}\mathrm{Tr}\left[\frac{\partial L_{ji,\theta_{1}}}{\partial\theta_{1}}\rho\frac{\partial L_{ji,\theta_{2}}^{\dagger}}{\partial\theta_{1}}\right]_{\theta_{1}=\theta_{2}=0}=\frac{1}{4}\sum_{i\ne j}\left[\sqrt{\eta_{ji}\rho_{ii}^{\mathrm{ss}}}-\sqrt{\eta_{ij}\rho_{jj}^{\mathrm{ss}}}\right]^{2}.
\end{align*}By using the inequality
\[
(a-b)^{2}\le\frac{1}{2}(a^{2}-b^{2})\ln\frac{a}{b}\;\;\;(a>0,b>0),
\]$\mathcal{X}$ is upper bounded by
\begin{equation}
\mathcal{X}\le\frac{\Sigma}{8},\label{eq:X_upper_bound}
\end{equation}where
\begin{equation}
\Sigma\equiv\sum_{i\ne j}\eta_{ji}\rho_{ii}^{\mathrm{ss}}\ln\frac{\eta_{ji}\rho_{ii}^{\mathrm{ss}}}{\eta_{ij}\rho_{jj}^{\mathrm{ss}}}.
\label{eq:Sigma_def}
\end{equation}$\Sigma$ corresponds to entropy production in classical stochastic thermodynamics. 

Next, we calculate $\mathcal{Z}_{1}$ (and $\mathcal{Z}_{2}$). We focus on $A_{j}(\bm{\theta}_{0})u(\bm{\theta}_{0})$ in Eqs.~\eqref{eq:Z1_def} and \eqref{eq:Z2_def}, which is used to calculate
\begin{align*}
A_{1}(\bm{\theta}_{0})u(\bm{\theta}_{0})&\propto\left.\frac{\partial\widetilde{\mathcal{L}}_{\theta_{1},\theta_{2}}}{\partial\theta_{1}}\left(\rho^{\mathrm{ss}}\right)\right|_{\theta_{1}=\theta_{2}=0}\\
&=\frac{1}{2}\sum_{i\ne j}\eta_{ji}\left(1-\sqrt{\frac{\eta_{ij}\rho_{jj}^{\mathrm{ss}}}{\eta_{ji}\rho_{ii}^{\mathrm{ss}}}}\right)\ket{\epsilon_{j}}\bra{\epsilon_{i}}\rho^{\mathrm{ss}}\ket{\epsilon_{i}}\bra{\epsilon_{j}}-\frac{1}{2}\sum_{i\ne j}\eta_{ji}\left(1-\sqrt{\frac{\eta_{ij}\rho_{jj}^{\mathrm{ss}}}{\eta_{ji}\rho_{ii}^{\mathrm{ss}}}}\right)\ket{\epsilon_{i}}\bra{\epsilon_{i}}\rho^{\mathrm{ss}}\\
&=\frac{1}{2}\sum_{i\ne j}\eta_{ji}\left(1-\sqrt{\frac{\eta_{ij}\rho_{jj}^{\mathrm{ss}}}{\eta_{ji}\rho_{ii}^{\mathrm{ss}}}}\right)\left(\rho_{ii}^{\mathrm{ss}}\ket{\epsilon_{j}}\bra{\epsilon_{j}}-\ket{\epsilon_{i}}\bra{\epsilon_{i}}\rho^{\mathrm{ss}}\right).
\end{align*}We substitute $\rho^{\mathrm{ss}}=\sum_{i}\rho_{ii}^{\mathrm{ss}}\ket{\epsilon_{i}}\bra{\epsilon_{i}}$ to obtain
\begin{align}
A_{1}(\bm{\theta}_{0})u(\bm{\theta}_{0})&\propto\sum_{i\ne j}\eta_{ji}\left(1-\sqrt{\frac{\eta_{ij}\rho_{jj}^{\mathrm{ss}}}{\eta_{ji}\rho_{ii}^{\mathrm{ss}}}}\right)\left(\rho_{ii}^{\mathrm{ss}}\ket{\epsilon_{j}}\bra{\epsilon_{j}}-\rho_{ii}^{\mathrm{ss}}\ket{\epsilon_{i}}\bra{\epsilon_{i}}\right)\nonumber\\&=\sum_{i\ne j}\eta_{ji}\left(\rho_{ii}^{\mathrm{ss}}\ket{\epsilon_{j}}\bra{\epsilon_{j}}-\rho_{ii}^{\mathrm{ss}}\ket{\epsilon_{i}}\bra{\epsilon_{i}}\right)\nonumber\\&=0,
 \label{eq:Z_zero}
\end{align}where the second line is identical to Eq.~\eqref{eq:Z_zero_DA}. From Eqs.~\eqref{eq:X_upper_bound} and \eqref{eq:Z_zero}, $\mathcal{Z}_1=\mathcal{Z}_2=0$ and hence we obtain
\begin{equation}
\frac{\mathrm{Var}\left[\Theta\right]}{\mathbb{E}\left[\Theta\right]^{2}}\ge\frac{2h^{\prime}(0)^{2}}{T\Sigma}.
\label{eq:QTUR2}
\end{equation}Equation~\eqref{eq:QTUR2} is the second result in the main text [Eq.~\QTURII{}]. 

\section{Continuous measurements}

\subsection{Jump measurement}
We introduce a jump measurement. The Kraus operator $\mathcal{V}_m$ for the jump measurement is given by
\begin{align}
\mathcal{V}_{0}&=\mathbb{I}-i\left(H-\frac{i}{2}\sum_{c}L_{c}^{\dagger}L_{c}\right)dt,\label{eq:V0_jump}\\
\mathcal{V}_{c}&=L_{c}\sqrt{dt}\;\;\;(c\ge 1).\label{eq:V1_jump}
\end{align}Equations~\eqref{eq:V0_jump} and \eqref{eq:V1_jump} satisfy the completeness relation:
\begin{align*}
\mathcal{V}_0^\dagger \mathcal{V}_0 + \sum_{c\ge 1}\mathcal{V}_{c}^{\dagger}\mathcal{V}_{c}  & =\mathbb{I}.
\end{align*}
\subsection{Diffusion measurement}

We introduce a diffusion measurement following Ref.~\cite{Gammelmark:2013:Bayes}. For simplicity, we consider a one-dimensional case, because its multidimensional generalization is straight forward. For diffusion measurement, the Kraus operator $\mathcal{V}_m$ is given by
\begin{align}
\mathcal{V}_{\Delta\mathcal{Y}}&=\sqrt{\mathscr{P}(\Delta\mathcal{Y})}\left[\mathbb{I}-iH\Delta t-\frac{1}{2}L^{\dagger}L\Delta t+L\Delta\mathcal{Y}\right],\label{eq:Kraus_diffusion}
\end{align}where $\Delta \mathcal{Y}$ is the output of the measurement and $\mathscr{P}(\cdot)$ is a Gaussian distribution with zero mean and variance $\Delta t$. Equation~\eqref{eq:Kraus_diffusion} satisfies the completeness relation upto $O(\Delta t)$:
\[
\int d\Delta \mathcal{Y}\,\mathcal{V}_{\Delta \mathcal{Y}}^{\dagger}\mathcal{V}_{\Delta \mathcal{Y}}=\mathbb{I}.
\]
Therefore, $\mathcal{V}_{\Delta\mathcal{Y}}$ constitutes a valid Kraus operator. Upto $O(\Delta \mathcal{Y})$ and $O(\Delta t^0)$, the probability of observing the output $\Delta \mathcal{Y}$ is
\begin{align}
p(\Delta\mathcal{Y})&=\mathrm{Tr}\left[\mathcal{V}_{\Delta\mathcal{Y}}\rho\mathcal{V}_{\Delta\mathcal{Y}}^{\dagger}\right]\nonumber\\&=\mathscr{P}(\Delta\mathcal{Y})\mathrm{Tr}\left[\rho+\left(L\rho+\rho L^{\dagger}\right)\Delta\mathcal{Y}\right].
 \label{eq:p_dY}
\end{align}From Eq.~\eqref{eq:p_dY}, upto $O(\Delta t)$, the mean and variance of $\Delta \mathcal{Y}$ are
\begin{align}
\mathbb{E}\left[\Delta\mathcal{Y}\right]&=\int d\Delta\mathcal{Y}\,p(\Delta\mathcal{Y})\Delta\mathcal{Y}=\mathrm{Tr}\left[L\rho+\rho L^{\dagger}\right]\Delta t,\label{eq:mean_dY}\\
\mathbb{E}\left[\Delta\mathcal{Y}^{2}\right]&=\int d\Delta\mathcal{Y}\,p(\Delta\mathcal{Y})\Delta\mathcal{Y}^{2}=\Delta t.\label{eq:variance_dY}
\end{align}This implies that the dynamics of $d\mathcal{Y}$ are given by the following Ito stochastic differential equation:
\begin{equation}
    d\mathcal{Y}=\mathrm{Tr}\left[L\rho+\rho L^{\dagger}\right]dt+dW,\label{eq:dYdt_def}
\end{equation}where $W$ is the standard Wiener process. Upto $O(\Delta t)$ contributions, the time-evolution of $\rho(t)$ given the output $\Delta \mathcal{Y}$ is
\begin{align}
\rho(t+\Delta t)&=\mathcal{V}_{\Delta\mathcal{Y}}\rho(t)\mathcal{V}_{\Delta\mathcal{Y}}^{\dagger}\nonumber\\&=\mathscr{P}(\Delta\mathcal{Y})\left[\rho-iH\rho\Delta t-\frac{1}{2}L^{\dagger}L\rho\Delta t+L\rho\Delta\mathcal{Y}+i\rho H\Delta t-\frac{1}{2}\rho L^{\dagger}L\Delta t+\rho L^{\dagger}\Delta\mathcal{Y}+L\rho L^{\dagger}\Delta\mathcal{Y}^{2}\right].\label{eq:rho_diffusion_def}
\end{align}In Eq.~\eqref{eq:rho_diffusion_def}, $d\mathcal{Y}^2=dt$, since $dW^2=dt$ for any non-anticipating functions \cite{Gardiner:2009:Book}. 
Let us introduce an unnormalized density operator $\tilde{\rho}$. For $\Delta t \to 0$, $\tilde{\rho}$ is governed by
\[
d\tilde{\rho}=\left(-i[H,\tilde{\rho}]+L\tilde{\rho}L^{\dagger}-\frac{1}{2}L^{\dagger}L\tilde{\rho}-\frac{1}{2}\tilde{\rho}L^{\dagger}L\right)dt+\left(L\tilde{\rho}+\tilde{\rho}L^{\dagger}\right)d\mathcal{Y}.
\]
The normalized density is given by $\rho=\tilde{\rho}/\mathrm{Tr}[\tilde{\rho}]$, which yields
\begin{align}
    d\rho&=\left(-i[H,\rho]-\frac{1}{2}L^{\dagger}L\rho-\frac{1}{2}\rho L^{\dagger}L+L\rho L^{\dagger}\right)dt+\left[L\rho+\rho L^{\dagger}-\rho\mathrm{Tr}\left(L\rho+\rho L^{\dagger}\right)\right]dW,\label{eq:QSD_rho}\\
    d\mathcal{Y}&=\mathrm{Tr}[L\rho+\rho L^{\dagger}]dt+dW.\label{eq:QSD_Y}
\end{align}Equations~\eqref{eq:QSD_rho} and \eqref{eq:QSD_Y} are known as quantum-state diffusion. 

\section{Examples}

\subsection{Two-level atom}

\begin{figure}
\begin{centering}
\includegraphics[width=13cm]{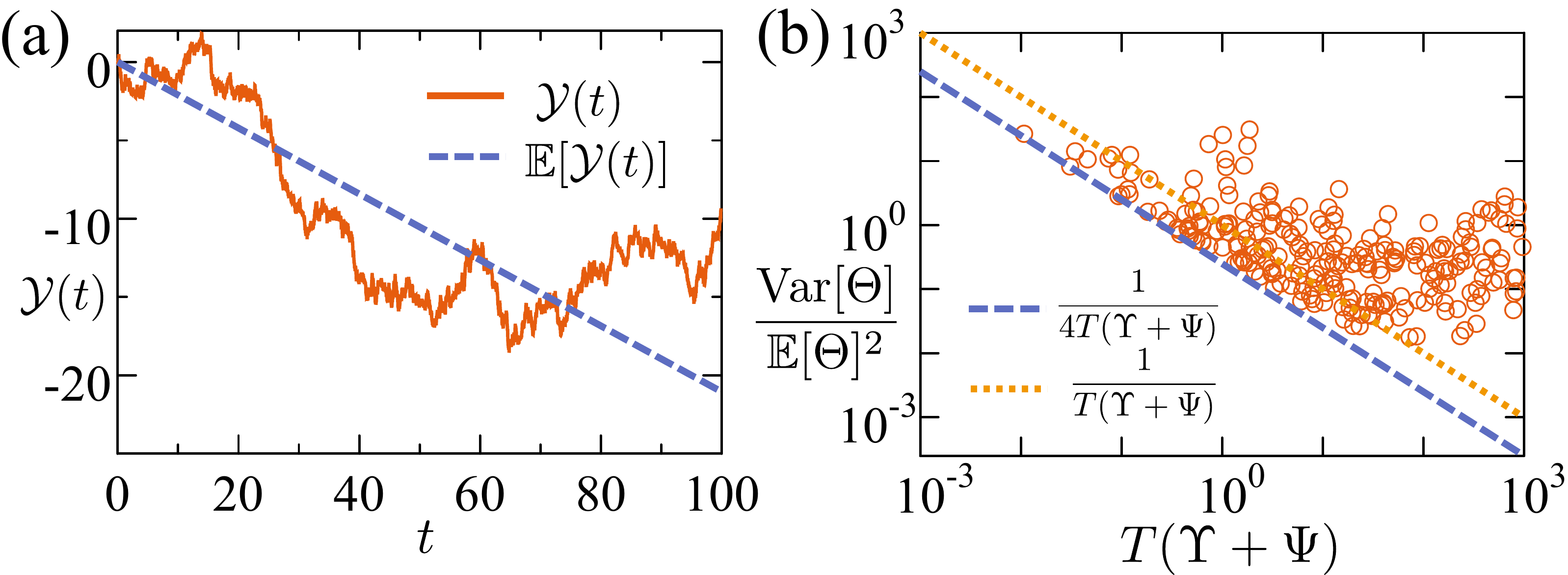}
\par\end{centering}
\caption{(a) Random trajectory of $\mathcal{Y}(t)$ (solid line) and its expectation
$\mathbb{E}[\mathcal{Y}(t)]$ (dashed line) as a function of $t$. Parameters
are $\Delta=1.0$, $\Omega=1.0$, and $\kappa=0.1$ (b) $\mathrm{Var}[\Theta_{\mathcal{Y}}]/\mathbb{E}[\Theta_{\mathcal{Y}}]$
(circles) for the diffusion measurement (homodyne detection)
as a function of $T(\Upsilon + \Psi)$.
Parameters are randomly selected from $\Delta\in[0.1,3.0]$,
$\Omega\in[0.1,3.0]$, $\kappa\in[0.1,3.0]$, and $T=100$. The dashed
and the dotted lines are $1/[4T(\Upsilon+\Psi)]$ and $1/[T(\Upsilon+\Psi)]$,
respectively. \label{fig:two_atom}}
\end{figure}

The Lindblad equation of the two-level atom is given by
\[
\frac{d\rho}{dt}=-i\left[H,\rho\right]+L\rho L^{\dagger}-\frac{1}{2}\left\{ L^{\dagger}L,\rho\right\},
\]where the Hamiltonian $H$ and the jump operator $L$ are defined by
\begin{align*}
H & =\Delta\ket{\epsilon_{e}}\bra{\epsilon_{e}}+\frac{\Omega}{2}\left(\ket{\epsilon_{e}}\bra{\epsilon_{g}}+\ket{\epsilon_{g}}\bra{\epsilon_{e}}\right),\\
L & =\sqrt{\kappa}\ket{\epsilon_{g}}\bra{\epsilon_{e}}.
\end{align*}Here, $\ket{\epsilon_e}$ and $\ket{\epsilon_g}$ are the excited and ground states, respectively; $\Delta$ is a detuning between the laser-field frequency and the atomic-transition frequency; $\Omega$ is the Rabi-oscillation frequency; and $\kappa$ is the decay rate. The steady-state density matrix is
\[
\rho^{\mathrm{ss}}=\left[\begin{array}{cc}
\rho_{gg}^{\mathrm{ss}} & \rho_{ge}^{\mathrm{ss}}\\
\rho_{eg}^{\mathrm{ss}} & \rho_{ee}^{\mathrm{ss}}
\end{array}\right]=\frac{1}{4\Delta^{2}+\kappa^{2}+2\Omega^{2}}\left[\begin{array}{cc}
4\Delta^{2}+\kappa^{2}+\Omega^{2} & -2\Delta\Omega+i\kappa\Omega\\
-2\Delta\Omega-i\kappa\Omega & \Omega^{2}
\end{array}\right],
\]
where $\rho_{ij}^{\mathrm{ss}}\equiv\braket{\epsilon_{i}|\rho^{\mathrm{ss}}|\epsilon_{j}}$. In the main text, we performed a computer simulation for the jump measurement. Here, we also carry out a computer simulation for the diffusion measurement. We first show a trajectory of $\mathcal{Y}(t)$ as a function of $t$. As indicated in the main text, from Eq.~\eqref{eq:QSD_Y}, it is given by
\begin{equation}
\mathcal{Y}(t)=\int_{0}^{t}d\mathcal{Y}=\int_{0}^{t}dt^{\prime} \mathrm{Tr}[L\rho+\rho L^{\dagger}]+\int_{0}^{t}dW,\label{Yt_def}
\end{equation}and its average is
\begin{equation}
\mathbb{E}[\mathcal{Y}(t)]=t \mathrm{Tr}[L\rho^{\mathrm{ss}}+\rho^{\mathrm{ss}}L^{\dagger}]=-\frac{4\sqrt{\kappa}\Delta\Omega}{4\Delta^{2}+\kappa^{2}+2\Omega^{2}}t.\label{eq:EY_def}
\end{equation}We plot an example of $\mathcal{Y}(t)$ as a function of $t$ in Fig.~\ref{fig:two_atom}(a), where the solid line is a random realization of $\mathcal{Y}(t)$ and the dashed line is its expectation $\mathbb{E}[\mathcal{Y}(t)]$, as shown by Eq.~\eqref{eq:EY_def}. 

We also check the QTUR for the diffusion measurement by randomly generating $\kappa$, $\Omega$, and $\Delta$ (the ranges of the parameters are shown in the caption of Fig.~\ref{fig:two_atom}(b)) and calculate $\mathrm{Var}[\Theta_{\mathcal{Y}}]/\mathbb{E}[\Theta_{\mathcal{Y}}]^{2}$, where $\Theta_\mathcal{Y}\equiv \int_0^T d\mathcal{Y}$. As mentioned in the main text, the lower bound of the diffusion measurement is $1/4$-times smaller than in the jump-measurement case, since $h^{\prime}(0)=1/2$. Figure~\ref{fig:two_atom}(b) plots $\mathrm{Var}[\Theta_{\mathcal{Y}}]/\mathbb{E}[\Theta_{\mathcal{Y}}]^{2}$ as a function of $\mathcal{I}_{Q}(0) = T(\Upsilon + \Psi)$ with circles. In Fig.~\ref{fig:two_atom}(b), we plot the lower bound of $1/(4\mathcal{I}_{Q}(0))$ with a dashed line. We also plot $1/(\mathcal{I}_{Q}(0))$, which is the lower bound of the jump-measurement case, with a dotted line. As can been seen, all realizations are located above $1/(4\mathcal{I}_{Q}(0))$, which verifies the QTUR of Eq.~\eqref{eq:QTUR1} [Eq.~\QTURI{} in the main text], while some realizations are below $1/(\mathcal{I}_{Q}(0))$, supporting the claim that the bound of the diffusion measurement is $1/4$-times smaller than the jump-measurement case.

\subsection{Three-level thermal machine}

\begin{figure}
\begin{centering}
\includegraphics[width=13cm]{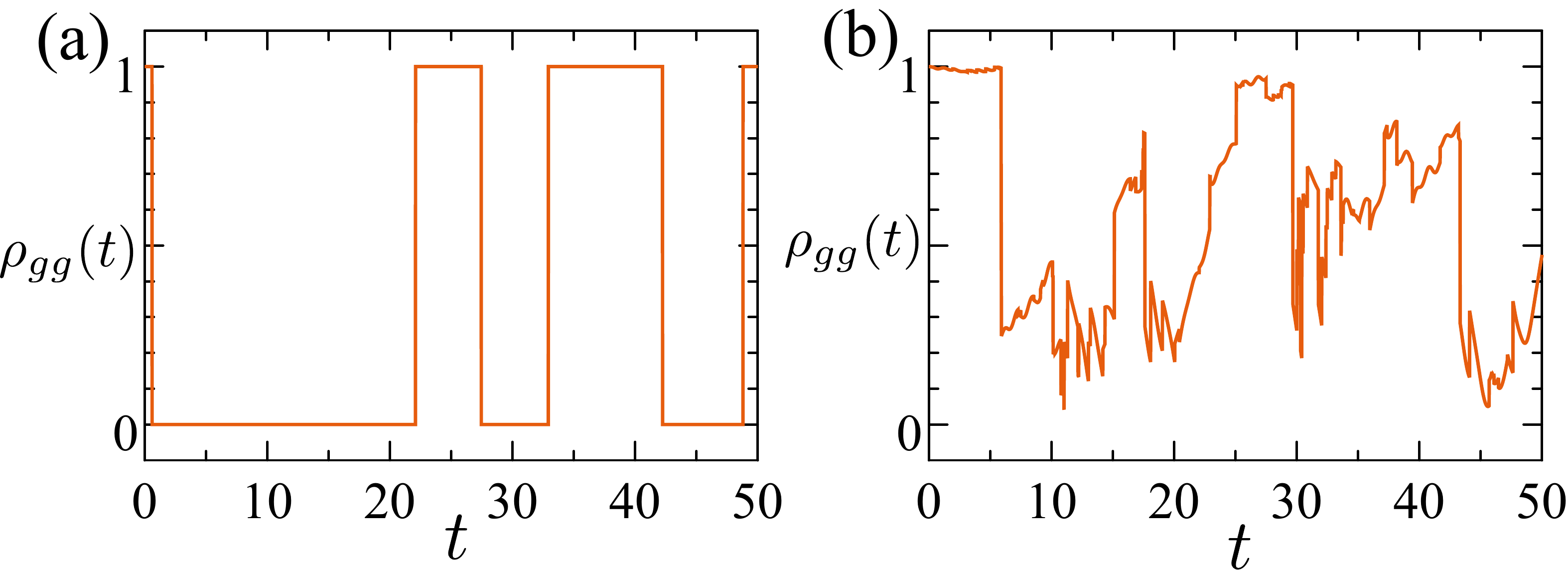}
\par\end{centering}
\caption{Trajectory of $\rho_{gg}(t)\equiv\braket{\epsilon_{g}|\rho(t)|\epsilon_{g}}$
for (a) standard jump measurement and (b) transformed jump measurement.
The trajectories are generated for $\beta_{1}=1.0$, $\beta_{2}=0.2$, $\beta_{3}=0.1$,
$\Omega_{1}=1.0$, $\Omega_{2}=5.0$, and $\text{\ensuremath{\gamma=0.1}}$.
The additional parameters for the transformed jump measurement are
$\zeta_{gA}=1.0$, $\zeta_{Ag}=0.2$, $\zeta_{AB}=-0.3$, $\zeta_{BA}=0.0$,
$\zeta_{gB}=0.1$, and $\zeta_{Bg}=-0.4$. \label{fig:Ref_qtraj_example}}
\end{figure}

We also consider a three-level thermal machine in the main text. We follow a description in Ref.~\cite{Paule:2018:PhDBook}. The Lindblad equation of the system is given by
\begin{equation}
\frac{d\rho}{dt}=-i\left[H,\rho\right]+\sum_{i\ne j}\left[L_{ji}\rho L_{ji}^{\dagger}-\frac{1}{2}\left\{ L_{ji}^{\dagger}L_{ji},\rho\right\} \right],\label{eq:Lindblad_Ref_def}
\end{equation}where the Hamiltonian $H$ and the jump operator $L_{ji}$ are defined by
\begin{align}
H & =\omega_3\ket{\epsilon_{B}}\bra{\epsilon_{B}}+\omega_{1}\ket{\epsilon_{A}}\bra{\epsilon_{A}},\label{eq:H_Ref}\\
L_{ji} & =\sqrt{\eta_{ji}}\ket{\epsilon_{j}}\bra{\epsilon_{i}}\;\;\;(i\ne j).\label{eq:Lij_Ref}
\end{align}Here, $\ket{\epsilon_g}$, $\ket{\epsilon_A}$, and $\ket{\epsilon_B}$ are energy levels, and $\eta_{ji}$ is the transition rate from $\ket{\epsilon_{i}}$ to $\ket{\epsilon_j}$. $\omega_1$, $\omega_2$, and $\omega_3 = \omega_1 + \omega_2$ are energy gaps between $\ket{\epsilon_A}\leftrightarrow \ket{\epsilon_g}$, $\ket{\epsilon_B}\leftrightarrow \ket{\epsilon_A}$, and $\ket{\epsilon_B}\leftrightarrow \ket{\epsilon_g}$, respectively. The transition rate fulfills the detailed balance condition:
\begin{align}
\frac{\eta_{gA}}{\eta_{Ag}} & =\frac{\gamma(n_{1}^{\mathrm{th}}+1)}{\gamma n_{1}^{\mathrm{th}}}=e^{\beta_{1}\omega_{1}},\label{eq:DB_Ag}\\
\frac{\eta_{AB}}{\eta_{BA}} & =\frac{\gamma(n_{2}^{\mathrm{th}}+1)}{\gamma n_{2}^{\mathrm{th}}}=e^{\beta_{2}\omega_{2}},\label{eq:DB_BA}\\
\frac{\eta_{gB}}{\eta_{Bg}} & =\frac{\gamma(n_{3}^{\mathrm{th}}+1)}{\gamma n_{3}^{\mathrm{th}}}=e^{\beta_{3}\omega_{3}},\label{eq:DB_Bg}
\end{align}where $n_{r}^{\mathrm{th}}\equiv(e^{\beta_{r}\omega_{r}}-1)^{-1}$, $\beta_r$ is the inverse temperature of $r$th heat reservoir, and $\gamma$ is the decay rate. The entropy production of the thermal machine is given by
\begin{equation}
\Phi=-\sum_{r=1}^{3}\beta_{r}\dot{Q}_{r},\label{eq:Sigma_TM_def}
\end{equation}where $\dot{Q}_{r}$ is the heat flux entering from the $r$th reservoir. $\dot{Q}_r$ is represented by
\begin{align}
\dot{Q}_{1} & =\mathrm{Tr}\left[\left(\mathcal{D}(\rho,L_{Ag})+\mathcal{D}(\rho,L_{gA})\right)H\right]=\omega_{1}\left(\eta_{Ag}\rho_{gg}^{\mathrm{ss}}-\eta_{gA}\rho_{AA}^{\mathrm{ss}}\right),\label{eq:Q1_def}\\
\dot{Q}_{2} & =\mathrm{Tr}\left[\left(\mathcal{D}(\rho,L_{BA})+\mathcal{D}(\rho,L_{AB})\right)H\right]=\omega_{2}\left(\eta_{BA}\rho_{AA}^{\mathrm{ss}}-\eta_{AB}\rho_{BB}^{\mathrm{ss}}\right),\label{eq:Q2_def}\\
\dot{Q}_{3} & =\mathrm{Tr}\left[\left(\mathcal{D}(\rho,L_{Bg})+\mathcal{D}(\rho,L_{gB})\right)H\right]=\omega_{3}\left(\eta_{Bg}\rho_{gg}^{\mathrm{ss}}-\eta_{gB}\rho_{BB}^{\mathrm{ss}}\right).\label{eq:Q3_def}
\end{align}Combining Eqs.~\eqref{eq:DB_Ag}--\eqref{eq:Q3_def}, we obtain
\begin{align*}
\Phi & =-\beta_{1}\omega_{1}\left(\eta_{Ag}\rho_{gg}^{\mathrm{ss}}-\eta_{gA}\rho_{AA}^{\mathrm{ss}}\right)-\beta_{2}\omega_{2}\left(\eta_{BA}\rho_{AA}^{\mathrm{ss}}-\eta_{AB}\rho_{BB}^{\mathrm{ss}}\right)-\beta_{3}\omega_{3}\left(\eta_{Bg}\rho_{gg}^{\mathrm{ss}}-\eta_{gB}\rho_{BB}^{\mathrm{ss}}\right)\\
 & =\sum_{i\ne j}\eta_{ji}\rho_{ii}^{\mathrm{ss}}\ln\frac{\eta_{ji}}{\eta_{ij}},
\end{align*}which is identical to the heat-dissipation rate in stochastic thermodynamics. Therefore, for the steady-state condition, we obtain $\Phi=\Sigma$. 

We first consider a standard jump measurement. The dynamics of the density matrix are given by the stochastic Schr{\"o}dinger equation  {[}Eq.~\SSEUrefrige{} in the main text{]}:
\begin{align}
d\rho=-i[H,\rho]dt+\sum_{i\ne j}\left(\rho\mathrm{Tr}\left[L_{ji}\rho L_{ji}^{\dagger}\right]-\frac{\left\{ L_{ji}^{\dagger}L_{ji},\rho\right\} }{2}\right)dt+\sum_{i\ne j}\left(\frac{L_{ji}\rho L_{ji}^{\dagger}}{\mathrm{Tr}[L_{ji}\rho L_{ji}^{\dagger}]}-\rho\right)d\mathcal{N}_{ji},\label{eq:SSE_Ref_def}
\end{align}where $d\mathcal{N}_{ji}$ is a noise increment. $d\mathcal{N}_{ji} = 1$ when the jump from $\ket{\epsilon_i}$ to $\ket{\epsilon_j}$ occurs at each time interval $dt$ and $d\mathcal{N}_{ji} = 0$ otherwise. Its expectation is given by $\mathbb{E}[d\mathcal{N}_{ji}]=\mathrm{Tr}[L_{ji}\rho^{\mathrm{ss}}L_{ji}^{\dagger}]dt=\eta_{ji}\rho_{ii}^{\mathrm{ss}}dt$. The corresponding Kraus operator $\mathcal{V}_m$ is given by
\begin{align}
\mathcal{V}_{ji}&=L_{ji}\sqrt{\Delta t}\;\;\;(i\ne j),\label{eq:Vij_Ref_def}\\
\mathcal{V}_{0} & =\mathbb{I}-i\left[H-\frac{i}{2}\sum_{i\ne j}L_{ji}^{\dagger}L_{ji}\right]\Delta t.\label{eq:V0_Ref_def}
\end{align}In the main text, we consider the observable
\begin{equation}
\Theta_{C}\equiv\sum_{i\ne j}R_{ji}\int_{0}^{T}d\mathcal{N}_{ji},\label{eq:Theta_C_def}
\end{equation}where $R_{ji}\in \mathbb{R}$ and $R_{ji}=-R_{ij}$ for all $i$ and $j$. Using Eq.~\eqref{eq:EP_scaling}, we find that the following relation holds:
\begin{align}
\mathbb{E}_{\theta}[d\mathcal{N}_{ji}]-\mathbb{E}_{\theta}[d\mathcal{N}_{ij}]&=\mathrm{Tr}\left[L_{ji,\theta}\rho^{\mathrm{ss}}L_{ji,\theta}^{\dagger}\right]dt-\mathrm{Tr}\left[L_{ij,\theta}\rho^{\mathrm{ss}}L_{ij,\theta}^{\dagger}\right]dt\nonumber\\&=\eta_{ji}\left[1+\theta\left(1-\sqrt{\frac{\eta_{ij}\rho_{jj}^{\mathrm{ss}}}{\eta_{ji}\rho_{ii}^{\mathrm{ss}}}}\right)\right]\rho_{ii}^{\mathrm{ss}}dt-\eta_{ij}\left[1+\theta\left(1-\sqrt{\frac{\eta_{ji}\rho_{ii}^{\mathrm{ss}}}{\eta_{ij}\rho_{jj}^{\mathrm{ss}}}}\right)\right]\rho_{jj}^{\mathrm{ss}}dt\nonumber\\&=(1+\theta)\left(\eta_{ji}\rho_{ii}^{\mathrm{ss}}-\eta_{ij}\rho_{jj}^{\mathrm{ss}}\right)dt\nonumber\\&=(1+\theta)\left[\mathbb{E}_{\theta=0}[d\mathcal{N}_{ji}]-\mathbb{E}_{\theta=0}[d\mathcal{N}_{ij}]\right].
 \label{eq:E_diff}
\end{align}Therefore, $\Theta_{C}$ satisfies the scaling condition of Eq.~\obsUscaling{} and thus the QTUR of Eq.~\eqref{eq:QTUR2} [Eq.~\QTURII{}] holds for $\Theta_{C}$. 

In the main text, we also consider a transformed jump measurement. First, note that the Lindblad equation is invariant under the following transformation:
\begin{align}
L_{ji}^{\prime} & =L_{ji}+\zeta_{ji}\mathbb{I},\label{eq:Lij_mod}\\
H^{\prime} & =H-\frac{i}{2}\sum_{i\ne j}\left[\zeta_{ji}^{*}L_{ji}-\zeta_{ji}L_{ji}^{\dagger}\right],\label{eq:H_mod}
\end{align}where $\zeta_{ji}\in \mathbb{C}$ is a parameter. $\zeta_{ji} = 0$ for all $i$ and $j$ recovers the standard jump measurement. The Lindblad equation remains unchanged when we replace $H$ and $L_{ji}$ with $H^\prime$ and $L_{ji}^\prime$, respectively. With this transformation, the stochastic Schr{\"o}dinger equation is 
\begin{align}
d\rho&=-i[H^{\prime},\rho]dt+\sum_{i\ne j}\left(\rho\mathrm{Tr}\left[L_{ji}^{\prime}\rho L_{ji}^{\prime\dagger}\right]-\frac{\left\{ L_{ji}^{\prime\dagger}L_{ji}^{\prime},\rho\right\} }{2}\right)dt+\sum_{i\ne j}\left(\frac{L_{ji}^{\prime}\rho L_{ji}^{\prime\dagger}}{\mathrm{Tr}[L_{ji}^{\prime}\rho L_{ji}^{\prime\dagger}]}-\rho\right)d\mathcal{N}_{ji}^{\prime},\label{eq:gen_SSE_def}
\end{align}where $d\mathcal{N}_{ji}^\prime$ is a noise increment whose meaning is identical to $d\mathcal{N}_{ji}$. The expectation of $d\mathcal{N}_{ji}^\prime$ is $\mathbb{E}[d\mathcal{N}_{ji}^\prime] = \mathrm{Tr}[L_{ji}^\prime \rho^\mathrm{ss} L_{ji}^{\prime \dagger}]dt$. From Eqs.~\eqref{eq:Lij_mod} and \eqref{eq:H_mod}, the corresponding Kraus operators are given by
\begin{align}
\mathcal{V}_{ji}^{\prime}&=L_{ji}^{\prime}\sqrt{\Delta t}\nonumber\\&=(L_{ji}+\zeta_{ji}\mathbb{I})\sqrt{\Delta t},\label{eq:Vijd_Ref_def}\\
\mathcal{V}_{0}^{\prime}&=\mathbb{I}-i\left[H^{\prime}-\frac{i}{2}\sum_{i\ne j}L_{ji}^{\prime\dagger}L_{ji}^{\prime}\right]\Delta t\nonumber\\&=\mathbb{I}-i\left[H-\frac{i}{2}\sum_{i\ne j}\left[\zeta_{ji}^{*}L_{ji}-\zeta_{ji}L_{ji}^{\dagger}\right]-\frac{i}{2}\sum_{i\ne j}(L_{ji}^{\dagger}+\zeta_{ji}^{*}\mathbb{I})(L_{ji}+\zeta_{ji}\mathbb{I})\right]\Delta t.\label{eq:V0d_Ref_def}
\end{align}We can easily show that Eqs.~\eqref{eq:Vijd_Ref_def} and \eqref{eq:V0d_Ref_def} satisfy $\mathcal{V}_{0}^{\prime\dagger}\mathcal{V}_{0}^{\prime}+\sum_{i\ne j}\mathcal{V}_{ji}^{\prime\dagger}\mathcal{V}_{ji}^{\prime}=\mathbb{I}$. 

As mentioned above, the Lindblad equation Eq.~\eqref{eq:Lindblad_Ref_def} remains unchanged under transformation by Eqs.~\eqref{eq:Lij_mod} and \eqref{eq:H_mod}. However, the quantum trajectory of the transformed stochastic Schrodinger equation becomes different from the standard jump-measurement case. The quantum trajectories of the standard jump measurement {[}Eq.~\eqref{eq:SSE_Ref_def}{]} are transitions between the energy eigenstates $\ket{\epsilon_{A}}$, $\ket{\epsilon_{B}}$, and $\ket{\epsilon_{g}}$ (Fig.~\ref{fig:Ref_qtraj_example}(a)). On the other hand, for the transformed jump measurement {[}Eq.~\eqref{eq:gen_SSE_def}{]}, quantum trajectories are not jumps between these energy eigenstates in general (Fig.~\ref{fig:Ref_qtraj_example}(b)). 

From Eq.~\eqref{eq:Lij_mod}, the mean noise increment $d\mathcal{N}_{ji}^\prime$ is
\begin{align}
\mathbb{E}[d\mathcal{N}_{ji}^{\prime}]&=\mathrm{Tr}\left[L_{ji}^{\prime}\rho^{\mathrm{ss}}L_{ji}^{\prime\dagger}\right]dt\nonumber\\
&=\left(\mathrm{Tr}\left[L_{ji}\rho^{\mathrm{ss}}L_{ji}^{\dagger}\right]+\mathrm{Tr}\left[L_{ji}\rho^{\mathrm{ss}}\zeta_{ji}^{*}\right]+\mathrm{Tr}\left[\zeta_{ji}\rho^{\mathrm{ss}}L_{ji}^{\dagger}\right]+\mathrm{Tr}\left[\zeta_{ji}\rho^{\mathrm{ss}}\zeta_{ji}^{*}\right]\right)dt\nonumber\\
&=\mathrm{Tr}\left[L_{ji}\rho^{\mathrm{ss}}L_{ji}^{\dagger}\right]dt+\left|\zeta_{ji}\right|^{2}dt\nonumber\\
&=\mathbb{E}[d\mathcal{N}_{ji}]+\left|\zeta_{ji}\right|^{2}dt.
\label{eq:E_dN_dash}
\end{align}In the second line, we used the fact that $\rho^\mathrm{ss}$ is diagonal in the energy eigenbasis. Again, we consider an observable $\Theta_C^\prime$, where $d\mathcal{N}_{ji}$ is replaced with $d\mathcal{N}_{ji}^\prime$:
\[
\Theta_{C}^{\prime}\equiv\sum_{i\ne j}R^\prime_{ji}\int_{0}^{T}d\mathcal{N}_{ji}^{\prime},
\]where $R_{ji}^\prime\in \mathbb{R}$ and $R_{ji}^\prime=-R_{ij}^\prime$ for all $i$ and $j$. Using Eq.~\eqref{eq:E_dN_dash}, we calculate Eq.~\eqref{eq:E_diff} for $d\mathcal{N}_{ji}^\prime$ as follows:
\begin{align*}
\mathbb{E}_{\theta}\left[d\mathcal{N}_{ji}^{\prime}\right]-\mathbb{E}_{\theta}\left[d\mathcal{N}_{ij}^{\prime}\right]&=\mathbb{E}_{\theta}\left[d\mathcal{N}_{ji}\right]-\mathbb{E}_{\theta}\left[d\mathcal{N}_{ij}\right]+\left|\zeta_{ji}\right|^{2}dt-\left|\zeta_{ij}\right|^{2}dt\\&=\left(1+\theta\right)\left(\mathbb{E}_{\theta=0}\left[d\mathcal{N}_{ji}\right]-\mathbb{E}_{\theta=0}\left[d\mathcal{N}_{ij}\right]\right)+\left|\zeta_{ji}\right|^{2}dt-\left|\zeta_{ij}\right|^{2}dt\\&=\left(1+\theta\right)\left(\mathbb{E}_{\theta=0}\left[d\mathcal{N}_{ji}^{\prime}\right]-\mathbb{E}_{\theta=0}\left[d\mathcal{N}_{ij}^{\prime}\right]-\left|\zeta_{ji}\right|^{2}dt+\left|\zeta_{ij}\right|^{2}dt\right)+\left|\zeta_{ji}\right|^{2}dt-\left|\zeta_{ij}\right|^{2}dt.
\end{align*}Therefore, when $\left|\zeta_{ji}\right|=\left|\zeta_{ij}\right|$ for all $i$ and $j$, the observable $\Theta_{C}^{\prime}$ satisfies the scaling condition and thus obeys the QTUR of Eq.~\eqref{eq:QTUR2} [Eq.~\QTURII{}].